\documentclass[twocolumn]{aastex63}

\usepackage{graphicx}
\usepackage{amsmath}
\usepackage{float}
\usepackage{natbib}
\usepackage{tabularx}
\usepackage{bm}
\usepackage{appendix}
\usepackage{longtable}

\graphicspath{{./}{figures/}}

\shorttitle{Compact Object Modeling in 47 Tuc}
\shortauthors{Ye et al.}

\begin{document}

\title{Compact Object Modeling in the Globular Cluster 47 Tucanae}

\author[0000-0001-9582-881X]{Claire S. Ye}
\affil{Department of Physics \& Astronomy, Northwestern University, Evanston, IL 60208, USA}
\affil{Center for Interdisciplinary Exploration \& Research in Astrophysics (CIERA), Northwestern University, Evanston, IL 60208, USA}
\correspondingauthor{Claire S.~Ye}
\email{shiye2015@u.northwestern.edu}

\author[0000-0002-4086-3180]{Kyle Kremer}
\affil{TAPIR, California Institute of Technology, Pasadena, CA 91125, USA}
\affil{The Observatories of the Carnegie Institution for Science, Pasadena, CA 91101, USA}

\author[0000-0003-4175-8881]{Carl L.~Rodriguez}
\affil{McWilliams Center for Cosmology and Department of Physics, Carnegie Mellon University, Pittsburgh, PA 15213, USA}

\author[0000-0002-1884-3992]{Nicholas Z.~Rui}
\affil{TAPIR, California Institute of Technology, Pasadena, CA 91125, USA}

\author[0000-0002-9660-9085]{Newlin C.~Weatherford}
\affil{Department of Physics \& Astronomy, Northwestern University, Evanston, IL 60208, USA}
\affil{Center for Interdisciplinary Exploration \& Research in Astrophysics (CIERA), Northwestern University, Evanston, IL 60208, USA}

\author[0000-0002-3680-2684]{Sourav Chatterjee}
\affil{Tata Institute of Fundamental Research, Homi Bhabha Road, Mumbai 400005, India}

\author[0000-0002-7330-027X]{Giacomo Fragione}
\affil{Department of Physics \& Astronomy, Northwestern University, Evanston, IL 60208, USA}
\affil{Center for Interdisciplinary Exploration \& Research in Astrophysics (CIERA), Northwestern University, Evanston, IL 60208, USA}

\author[0000-0002-7132-418X]{Frederic A.~Rasio}
\affil{Department of Physics \& Astronomy, Northwestern University, Evanston, IL 60208, USA}
\affil{Center for Interdisciplinary Exploration \& Research in Astrophysics (CIERA), Northwestern University, Evanston, IL 60208, USA}

\begin{abstract}
The globular cluster 47~Tucanae (47~Tuc) is one of the most massive star clusters in the Milky Way and is exceptionally rich in exotic stellar populations. For several decades it has been a favorite target of observers, and yet it is computationally very challenging to model because of its large number of stars ($N\gtrsim 10^6$) and high density. Here we present detailed and self-consistent 47~Tuc models computed with the \texttt{Cluster Monte Carlo} code (\texttt{CMC}). The models include all relevant dynamical interactions coupled to stellar and binary evolution, and reproduce various observations, including the surface brightness and velocity dispersion profiles, pulsar accelerations, and numbers of compact objects. We show that the present properties of 47~Tuc are best reproduced by adopting an initial stellar mass function that is both bottom-heavy and top-light relative to standard assumptions \citep[as in, e.g.,][]{Kroupa2001}, and an initial Elson profile \citep{Elson1987} that is overfilling the cluster's tidal radius. We include new prescriptions in \texttt{CMC} for the formation of binaries through giant star collisions and tidal captures, and we show that these mechanisms play a crucial role in the formation of neutron star binaries and millisecond pulsars in 47~Tuc; our best-fit model contains $\sim 50$ millisecond pulsars, $70\%$ of which are formed through giant collisions and tidal captures. Our models also suggest that 47~Tuc presently contains up to $\sim 200$ stellar-mass black holes, $\sim 5$ binary black holes, $\sim 15$ low-mass X-ray binaries, and $\sim 300$ cataclysmic variables.
\end{abstract}


\section{Introduction}\label{sec:intro}
As laboratories for studying both small-scale stellar physics and gravitational dynamics as well as larger-scale galaxy formation and evolution, globular clusters (GCs) have been a main focus of computational $N$-body modeling for several decades. They are efficient in producing exotic objects such as X-ray binaries \citep[e.g.,][]{Heinke+2005,Clark_1975}, millisecond pulsars \citep[MSPs; e.g.,][]{Ransom_2008}, and cataclysmic variables (CVs; \citealp[e.g.,][]{Knigge_2012}) given their high stellar densities. GCs may also account for a significant fraction of the overall binary black hole (BH) merger rate in the local universe \citep[][and references therein]{Abbott+2021,Kremer+2020catalog,Rodriguez_2021rnaas}. Furthermore, their masses, sizes, and spatial distributions provide unique probes into the formation of their host galaxies \citep[e.g.,][]{Choski+2018}.

Among Galactic GCs, 47~Tucanae (47~Tuc) is particularly interesting. It is one of the most dense, massive, and luminous GCs in the Milky Way \citep[][2010 edition]{Harris_1996} and contains large numbers of exotic sources, including $28$ MSPs \citep{psr_catalog}, $6$ low-mass X-ray binaries (LMXBs; \citealp{MillerJones_2015,Heinke+2005}), and $43$ CVs \citep{RSandoval_2018} observed to date. Since it is one of the closest GCs to Earth and is located at high Galactic latitude \citep[][2010 edition]{Harris_1996}, there are plenty of observational data on 47~Tuc, many of them from the Hubble Space Telescope (HST), Chandra, and more recently also from Gaia. For example, shortly after the launch of HST, \citet{Guhathakurta_1992_47Tuc} exploited its ability to resolve individual stars in crowded fields, constructed an accurate color-magnitude diagram (CMD), and placed an upper limit on the mass of compact objects at the cluster center. Subsequent HST campaigns aimed at obtaining detailed photometry of the cluster core \citep{deMarchi_1993_47tuc} and deriving stellar luminosity and mass functions \citep{Santiago_1996_47tuc}. CV candidates in the cluster were found through many epochs of observation, long exposure times and almost continuous observing \citep{Edmonds_1996_47tuc, Shara_1996_47tuc,Gilliland_1995_47tuc}. Combining X-ray observations from Chandra and $120$ orbits of HST, \citet{Edmonds_2003a,Edmonds_2003b} observed the largest number of optical counterparts to the X-ray sources in any GC, and detected many CVs, active binaries, and constrained the total number of MSPs. More recently, to further explore the dynamical structure of 47~Tuc, several studies obtained proper motions and line-of-sight velocities with HST \citep[e.g.,][]{McLaughlin_2006,Watkins+2015,Heyl_2017}, Gaia \citep{Baumgardt+2019,Vasiliev+2021}, MUSE \citep{Kamann+2018}, and archival/literature data from a combination of instruments \citep{Baumgardt_2017, Baumgardt_Hilker2018} to construct velocity distributions and examined the rotation and velocity anisotropy of the cluster \citep[e.g.,][]{Bellini_2017}.

Extensive previous studies have modeled 47~Tuc using static models, including single-mass King models \citep[e.g.,][]{Illingworth_Illingworh_1976}, multi-mass generalizations of King models \citep[e.g.,][]{DaCosta_Freeman_1985}, Michie-King models \citep[e.g.,][]{Meylan_1988,Meylan_1989}, models with added rotations \citep[e.g.,][]{Davoust_1986,Bellini_2017}, and models incorporating the effects of the Galactic tidal field on the structure of the cluster \citep[e.g.,][]{Henault-Brunet+2019}. However, there have been few previous attempts at modeling the cluster using the $N$-body method, despite this wealth of observational data on 47~Tuc. This is not surprising since modeling GCs as massive and dense as 47~Tuc requires significant computational time and sophisticated $N$-body codes. It takes more than a year for the current best direct $N$-body code, NBODY6++GPU \citep{Wang+2015nbody6}, to simulate a massive GC similar to 47~Tuc with $\sim 10^6$ stars. Therefore, the more rapid Monte Carlo technique is necessary to explore the parameter space of initial conditions for 47~Tuc (see Section~\ref{sec:method}). The only two previous studies of 47~Tuc using $N$-body modeling either rely upon scaling-up of low mass models \citep{Baumgardt_Hilker2018}, or apply artificially large natal kicks for BHs \citep{Giersz47Tuc2011} so there were only a few BHs in their models. Models without a realistic number of BHs may still show close matches with observations such as the surface brightness profile \citep{Giersz47Tuc2011}, but would potentially miss the BH dynamics that also significantly affects other exotic objects through gravitational encounters. To uncover the cluster's dynamical properties and understand all evolutionary pathways that shape the observable properties of 47~Tuc and other similar clusters, we develop accurate and self-consistent models of 47~Tuc that will match different observed cluster features simultaneously. Significantly, these models include most up-to-date prescriptions for compact object formation, which in particular enables us to examine pulsar and BH dynamics in 47~Tuc for the first time.

The \texttt{CMC} Cluster Catalog models in \citet{Kremer+2020catalog} are representative of most typical present-day Milky Way GCs, so we first searched for a best-fit 47~Tuc model in the catalog using the $\chi^2$ statistic method \citep{Rui+matching2021}. We found that while some catalog models can match the inner parts of the observed surface brightness and velocity dispersion profiles, there are large discrepancies at the outer parts of the good-fit models (Figure~\ref{fig:maingrid_massive}). This illustrates the necessity to simulate custom models for 47~Tuc with smaller core radii (less compact objects, especially BHs) and more low-mass stars to puff up the outer part of the cluster.

This paper is organized as follows. In Section~\ref{sec:method}, we summarize the main computational methods used to model 47~Tuc, and describe the implementation of tidal capture and collision with giant stars to our cluster simulations. We demonstrate that our 47~Tuc models fit the observations by comparing to multiple observational properties such as the surface brightness profile and velocity dispersion profile in Section~\ref{sec:match}. We show the compact object populations in a best-fit model in Section~\ref{sec:compact}. In Section~\ref{sec:discuss}, we discuss models constructed in search of the initial conditions that best agree with the observational data, caveats and uncertainties in our models. Finally, in Section~\ref{sec:conclusion}, we summarize our results.

\section{Methods}\label{sec:method}
We simulate 47~Tuc using the \texttt{Cluster Monte Carlo} code (\texttt{CMC}; \citealp[e.g.,][and references therein]{Rodriguez+2021CMC}), a public, fully parallelized Monte Carlo code based on classic work by \citet{henon1971monte, henon1971montecluster}. It is capable of modeling the complete dynamical evolution of realistic dense star clusters containing $\sim10^6$ stars over a Hubble time in just a few days. It allows us to compute self-consistent models of even the largest star clusters, covering the full range of masses all the way to the brightest systems. \texttt{CMC} incorporates all relevant physics, including two-body relaxation, tidal mass loss, and strong dynamical interactions of single stars and binaries (e.g., exchange interactions, tidal capture, direct collisions). Stellar evolution is fully coupled to the stellar dynamics and is computed with the publicly available \texttt{COSMIC} software \citep{cosmic} based on \citet{hurley2000comprehensive,hurley2002evolution}. The \texttt{Fewbody} package is used to directly integrate all three- and four-body gravitational encounters \citep{fregeau2004stellar,fregeau2007monte}, including post Newtonian dynamics for BHs \citep{antognini2014rapid,amaro2016relativistic,Rodriguez+repeated2018,rodriguez2018postb}. In the following subsections, we describe simulation features adopted here that differ from our past \texttt{CMC} assumptions.

\subsection{Tidal Capture}\label{subsec:tc}
Following the early works of \citet{Fabian+1975}, \citet{Press_Teukolsky_1977}, and \citet{Lee_Ostriker_1986}, we implemented tidal capture of MS stars into \texttt{CMC}. When a compact object and a MS star or two MS stars fly close during single-single encounters, tidal force will perturb the stars and excite non-radial oscillation within them. The oscillation energy comes from the two objects' kinetic energy, and the stars are captured if the oscillation energy exceeds that of their initial kinetic energy. The amount of energy deposited in a star from one passage of encounter can be written as 
\begin{equation}
    \Delta E_{osc} = \frac{GM_*^2}{R_*} \left(\frac{M}{M_* }\right)^2 \sum^\infty_{l=2}\left(\frac{R_*}{R_{p}}\right)^{2l+2} T_l(\eta_*),
\end{equation}
where $M_*$ and $R_*$ are the mass and the radius of the star, and $M$ is the mass of a compact object or another MS star. $R_{p}$ is the distance of closest approach. $T_l$ is a dimensionless function measuring energy contribution from different harmonic modes $l$. $\eta_*$ is a measure of the duration of periastron passage, and is written as 
\begin{equation}
    \eta_* = \left(\frac{M_*}{M_*+M}\right)^{1/2} \left(\frac{R_p}{R_*}\right)^{3/2}.
\end{equation}

We adopt the fitting formulae obtained from \citet{Lee_Ostriker_1986} in \citet{Kim_Lee_1999} to calculate the cross sections and pericenter distances $R_p$. The dimensionless function $T_l$ and the oscillation energy in the MS stars are estimated using the formulae in \citet{PZ+1993}, which were calculated by fitted to the hydrodynamical studies \citep{Lee_Ostriker_1986,Ray+1987}. We include only the contribution of the quadrupole and octupole terms in the tides, while higher-order corrections do not contribute significantly to the oscillation energy.

Since these prescriptions assumed polytropic stellar models, we apply polytropic index $n=1.5$ for low-mass MS stars with $M < 0.7~\rm{M_{\odot}}$ and naked helium MS stars, which correspond to star type 0 and 7 in \texttt{COSMIC} \citep{cosmic}, respectively. We use $n=3$ for other MS stars with $M > 0.7~\rm{M_{\odot}}$, corresponding to star type 1 \citep{McMillan+1987,Kim_Lee_1999}. For a comparison between the simple polytropic stellar models with detailed stellar models, see Appendix~\ref{app:tccomparison}.

The initial total energy of the two stars at infinity can be written as $E_{orb} = \frac{1}{2} \mu v_{\infty}^2$, where $\mu$ is the reduced mass and $v_\infty$ is the relative velocity at infinity. We assume that if the total oscillation energy deposited in the stars \textit{during the first passage} is larger than the initial total energy, then the two stars are bound and form a tidal capture binary. The binary semi-major axis immediately after tidal capture is set to be $2 R_p$, assuming angular momentum conservation during tidal capture and immediate circularization of the binary after tidal capture. We use this simple treatment of final binary separations to maximize the number of binaries formed from tidal capture \citep{Lee_Ostriker_1986}, especially NS--MS star binaries that could become redback MSPs in clusters (see also Section~\ref{subsec:pulsar}). In the special cases when Roche lobe overflow starts at $2 R_p$, we merge the two interacting objects instead of forming a binary.

\subsection{Giant Star Collision}\label{subsec:giant}
We implemented special treatments for single-single collisions involving giant stars into \texttt{CMC}. Here we define ``collision" as when the two interacting objects pass within a small factor of the sum of their radii or the radius of a giant star. When giant star collision is activated in \texttt{CMC}, we assume that a giant star collides with another star or compact object when $R_p \leq 1.3 R_g$ following \citet{Lombardi+2006}, where $R_p$ is the pericenter distance between the two objects and $R_g$ is the radius of the giant star. For the collision of two giants, we set the collision pericenter distance limit to $R_p \leq 1.3(R_{g1} + R_{g2})$, where $R_{g1}$ and $R_{g2}$ are the radii of the giants. At the same time, the ``sticky sphere" direct collision prescription ($R_p < R_1 + R_2$, where $R_1$ and $R_2$ are the radii) is applied to all other stars, where the mass of the collision product is equal to the total mass of the two colliding objects \citep{fregeau2007monte}.

These collisions with giant stars can lead to the formation of tight binaries. Qualitatively, we expect a collision involving at least one giant star to proceed similar to the common envelope process where the cores orbit inside an extended envelope. Through drag forces, the binary inspirals and deposits energy into the envelope until eventually the envelope is expelled, leaving behind a compact binary. We assume that there is no mass transfer to a companion star from the envelopes of the giant stars during these collisions. For single-single collisions involving only one giant, we assume that a binary star formed consists of the core of the giant star and the interacting main-sequence star (MS star) or compact object.

To calculate the binary semi-major axis, we adopt equations (3)-(5) below -- equations (4)-(6) from \citet{Ivanova_2006} -- based on smoothed particle hydrodynamics simulations \citep{Lombardi+2006}. The semi-major axis is chosen as the minimum of $a_{coll}$ and $a_{ce}$ from these equations.

$a_{coll}$ is calculated as
\begin{equation}
    a_{coll} = \frac{R_p}{3.3 (1-e_{coll}^2)},
\end{equation}

\noindent where
\begin{equation}
   e_{coll} = 0.88 - \frac{R_p}{3 R_g}.
\end{equation}

\noindent $a_{ce}$ can be obtained by solving the equation
\begin{equation}
    \frac{(M_g+M)v_{\infty}^2}{2} + \alpha_{ce}\frac{G M_c M}{2a_{ce}} = \frac{G M_g (M_g-M_c)}{\lambda R_g},
\end{equation}
with $e_{ce} = 0$. $M_g$, $M$, and $M_c$ are the giant's mass, the mass of the non-giant object, and the core mass of the giant, respectively. $\alpha_{ce} = 1$ and $\lambda = 0.5$ are parameters related to common-envelope evolution (\citealp[e.g.,][]{hurley2002evolution}; see also discussions in Appendix~\ref{app:tcgc_uncer}).

For collisions between two giant stars, we instead calculate the binary semi-major axis using the expression
\begin{multline}
    \frac{(M_{g1}+M_{g2})v_{\infty}^2}{2} + \alpha_{ce}\frac{G M_{c1} M_{c2}}{2a_{ce}} \\
    = \frac{G M_{g1} (M_{g1}-M_{c1})}{\lambda R_{g1}}+\frac{G M_{g2} (M_{g2}-M_{c2})}{\lambda R_{g2}}
\end{multline}
with $e_{ce} = 0$. $M_{g1}$, $M_{c1}$, $M_{g2}$, and $M_{c2}$ are the masses and the core masses of the first and second giants, respectively. The final binary is composed of the cores of the two giants and both giant envelopes are ejected.

In special cases where the MS star or a giant core starts mass transfer after the binary semi-major axis is computed, we assume the collision does not lead to binary formation. Instead we merge the two original interacting objects to avoid Roche Lobe overflow during a common-envelope phase, the outcome of which is highly uncertain. The stellar types of the merger products follow those in Table~2 of \citet{hurley2002evolution}.

\subsection{47 Tuc Simulations}\label{subsec:models}
For 47~Tuc simulations, we fix the metallicity $Z = 0.0038$ \citep[][2010 edition]{Harris_1996}, the Galactocentric distance $r_{gc} = 7.4~$kpc \citep{Harris2010catalog, Baumgardt+2019} and the initial binary fraction $f_b = 0.022$ \citep{Giersz47Tuc2011}. Binary companion masses are drawn from a flat distribution in mass ratio $q$, in the range $0.1-1$ \citep[e.g.,][]{Duquennoy_Mayor_1991}. Binary orbital periods are drawn from a distribution flat in log scale from Roche Lobe overflow to the hard/soft boundary, and binary eccentricities are drawn from a thermal distribution \citep[e.g.,][]{Heggie1975}. To calculate the tidal radius of a cluster, we assume a circular orbit with radius $r_{gc}$ around the Galactic center and a flat rotation curve of the galaxy with circular velocity $v_{gc} = 220~\rm{km\,s^{-1}}$. The tidal radius of a cluster is given by
\begin{equation}
    r_t = \left(\frac{GM_{clu}}{2v_{gc}^2}\right)^\frac{1}{3}r_{gc}^\frac{2}{3},
\end{equation}
where $M_{clu}$ is the total mass of the cluster \citep{Spitzer_1987,Chatterjee_2010}. This calculation of tidal radius is valid for 47~Tuc since the cluster's orbit in the galaxy has a small eccentricity ($<0.2$) and is almost circular \citep{Vasiliev+2021}.

Following \citet{Giersz47Tuc2011}, we adopt a two-part power-law initial mass function (IMF) with masses in the range of $0.08-150~\rm{M_{\odot}}$, a break mass at $0.8~\rm{M_{\odot}}$ and power-law slopes of $\alpha_1 = 0.4$ and $\alpha_2 = 2.8$ for the lower and higher mass part, respectively. Neutron stars (NSs) formed from core-collapse supernovae receive natal kicks drawn from a Maxwellian distribution with a standard velocity deviation $\sigma_{\rm{NS}} = 265~\rm{km\, s^{-1}}$ \citep{Hobbs+2005}. The natal kicks for NSs from electron-capture supernovae and accretion-induced collapse are also drawn from a Maxwellian distribution but with a smaller standard deviation $\sigma_{ecsn} = 20~\rm{km\,s^{-1}}$ \citep{Kiel+2008,Ye_msp_2019}. We apply mass fallback kicks for BHs and sample their natal kicks from the same Maxwellian distribution for core-collapse NSs but with scaled-down velocities \citep{Fryer+2012}. The kicks are reduced in magnitude by a fractional mass of fallback material: $v_{\rm{BH}} = v_{\rm{NS}}(1-f_{fb})$, where $v_{\rm{NS}}$ is the velocity of NS drawn from \citet{Hobbs+2005} for core-collapse supernovae and $f_{fb}$ is the fractional parameter of the stellar envelope that falls back upon core collapse \citep{Belczynski+2002,Fryer+2012,Morscher+2015}. Compared to \citet{Giersz47Tuc2011}, our simulations treat BHs and NSs more realistically by applying fallback BH natal kicks and allowing for NSs to form in electron-capture supernovae and accretion-induced collapses with low kicks, which are essential for explaining the large number of pulsars in GCs \citep[e.g.,][]{Pfahl+2002,Podsiadlowski+2004,Ye_msp_2019}.

We assume the 47~Tuc models are initially described by Elson density profiles \citep{Elson1987}, 
\begin{equation}
    \rho(r) = \rho_0 \left(1+\frac{r^2}{a^2}\right)^{-\frac{\gamma+1}{2}},
\end{equation}

\noindent where $\gamma$ is a free parameter of the power-law slope ($\gamma = 4$ gives a Plummer sphere; \citealp{Plummer_1911}), and $a$ is a scaling parameter.

Unlike previous \texttt{CMC} papers that varied the initial cluster parameter space in a grid format \citep[e.g.,][]{Kremer+2020catalog,Weatherford+2021}, here we have run $15$ simulations (see Table~\ref{tab:allsim} in Appendix~\ref{app:allsimulations}) exploring a specific parameter space for 47~Tuc in the initial number of stars, density profile, binary fraction, virial radius, tidal radius and IMF to search for a best-fit model where the model cluster's late time ($9-13.8\,\rm{Gyr}$) surface brightness profile (SBP) and velocity dispersion profile (VDP) first match the observations. We also vary $\gamma$ of the Elson profile to analytically fit the outer slope of the observed surface brightness profile. All simulations are evolved to $13.8~\rm{Gyr}$. Table~\ref{tab:IC} shows the initial conditions of our best-fit model for 47~Tuc.

\startlongtable
\begin{deluxetable*}{cccccccccccc}
\tabletypesize{\scriptsize}
\tablewidth{0pt}
\tablecaption{Initial Conditions for the Best-fit Model \label{tab:IC}}
\tablehead{
\colhead{$\rm{N}$} & \colhead{$\rm{\gamma}$} & \colhead{$\rm{r_v}$} & \colhead{$\rm{Z}$} & \colhead{$\rm{r_{gc}}$} & \colhead{$\rm{r_t}$} & \colhead{$\rm{M_{min}}$} & \colhead{$\rm{M_{br}}$} & \colhead{$\rm{M_{max}}$} & \colhead{$\rm{\alpha_1}$} & \colhead{$\rm{\alpha_2}$} & \colhead{$\rm{f_b}$}
}
\startdata
$3\times10^6$ & $2.1$ & $4$~pc & $0.0038$ & $7.4$~kpc & $182.17$~pc & $0.08~\rm{M_{\odot}}$ & $0.8~\rm{M_{\odot}}$ & $150~\rm{M_{\odot}}$ & $0.4$ & $2.8$ & $0.022$\\
\enddata
\tablecomments{From left to right: initial number of stars, Elson profile $\gamma$ index, initial virial radius, metallicity, Galactocentric distance, initial tidal radius, minimum mass, break mass, and maximum mass in the IMF, respectively, power-law slopes for the lower mass and higher mass part of the IMF, respectively, and initial binary fraction.}
\end{deluxetable*}

\section{Matching Observed Cluster Properties}\label{sec:match}
\begin{figure*}
\begin{center}
\includegraphics[width=\textwidth]{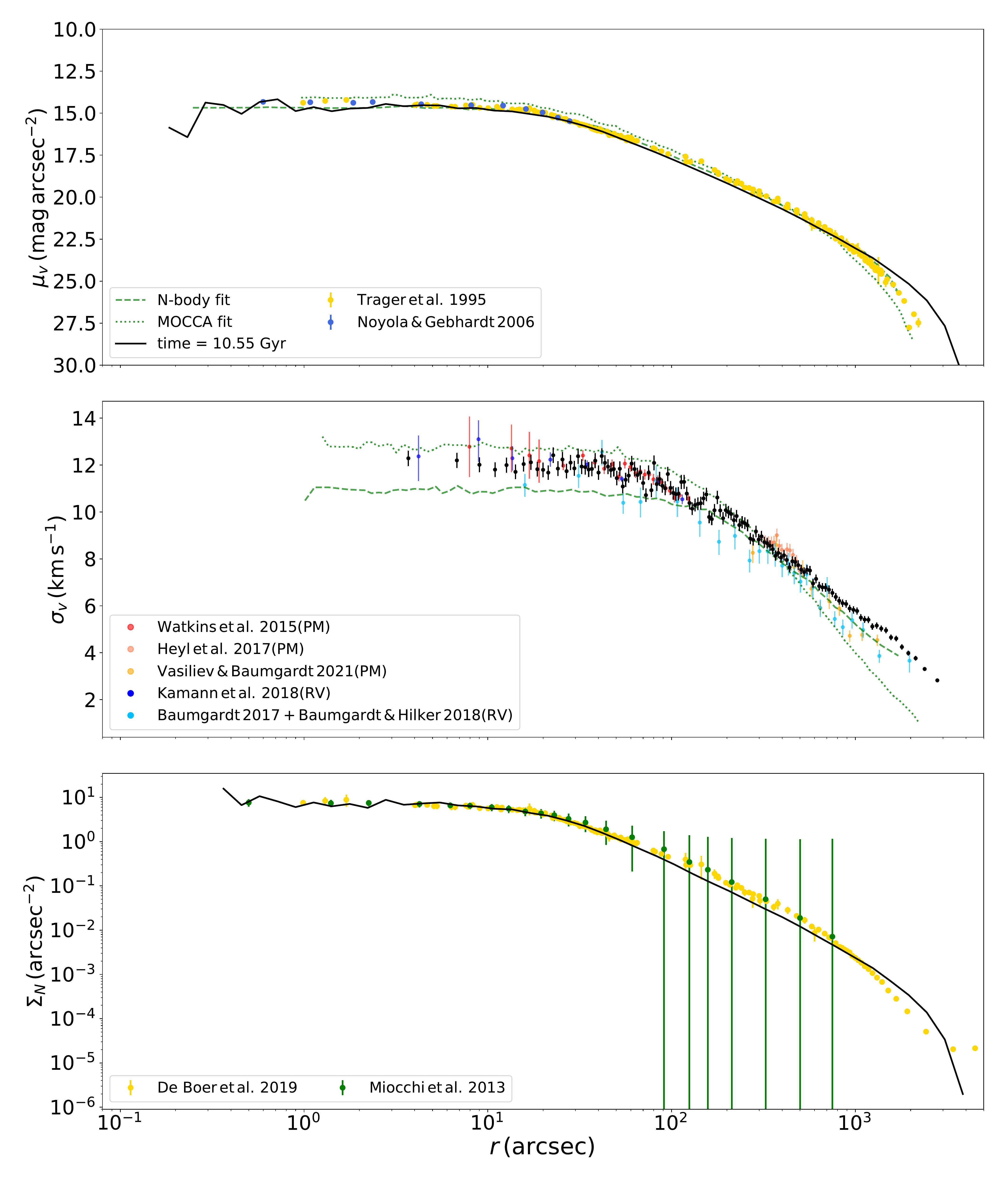}
\caption{Comparison between our model SBP, VDP and NDP with the observations. The model profiles are shown by the black curves and black markers. The observed SBPs are from \citet{Trager1995,Noyola_2006}, VDPs are from \citet{Watkins+2015,Heyl_2017,Kamann+2018,Baumgardt_2017,Baumgardt_Hilker2018,Vasiliev+2021} and NDPs are from \citet{Miocchi+2013,deboer+2019}. ``PM" stands for proper motion, and ``RV" stands for radial velocity. When available, fits from the MOCCA model (dotted curves; \citealp{Giersz47Tuc2011}) and the direct $N$-body model (dashed curves; \citealp{gc_catalog}) of 47~Tuc are also shown. Our model profiles agree well with the observations. \label{fig:sbp_vdp_ndp}}
\end{center}
\end{figure*}

We found best-fit models by comparing the $\chi^2$ likelihood between different snapshots (time steps) of the simulation and selecting the ones with small $\chi^2$ as in \citet{Rui+matching2021}. The model SBP is calculated by making two-dimensional projections at each time step of a simulation, assuming spherical symmetry \citep[also see][]{Kremer+2018_3201}. We exclude very bright stars with luminosity $L > 12~\rm{L_{\odot}}$ to avoid large fluctuations in the SBP, and also low-mass, faint stars with $M\leq0.85~\rm{M_{\odot}}$. Changing the cutoff luminosity from $12~\rm{L_{\odot}}$ to $15~\rm{L_{\odot}}$ has negligible effect on the SBP fit. We use the method in \citet{Pryor_Meylan_1993} to find the model VDP, including in the calculation all giant stars and upper MS stars with a cutoff mass $M>0.85~\rm{M_{\odot}}$ (the main-sequence turnoff mass of the model is $\sim 0.9~\rm{M_{\odot}}$). Small variations in the cutoff mass for the VDP calculations do not affect the fit. For more details on the SBP and VDP calculations, see \citet{Kremer+2018_3201} and \citet{Rui+matching2021}, and the references therein. Similar to the VDP, we use the same selection of stars to calculate the projected model number density profile (NDP). Note that, unlike the VDP, the model NDP is very sensitive to changes in the cutoff mass. The cluster properties only vary negligibly between $\sim 9-12~$Gyr, and Figure~\ref{fig:sbp_vdp_ndp} shows the snapshot at $10.55~$Gyr as a representative of the best-fit 47~Tuc models. At this time, the best-fit model is able to closely match the observed SBP, VDP,  NDP, and pulsar acceleration distribution simultaneously.

The initial tidal radius of the cluster with a circular orbit at $r_{gc}=7.4~$kpc is $182~$pc. The best-fit cluster starts out to $r_{\mathrm{max}}=800~$pc initially, and is overfilling its tidal radius ($r_{\mathrm{max}}$ is a free parameter in our simulations). Observationally, young star clusters (potential GC progenitors) appear to not be tidally truncated, and may be surrounded by unbound stars at the outer region that have not yet been stripped away by the Galactic tidal force \citep[e.g.,][]{Elson1987,Elson+1989,Mackey_Gilmore_2003}. Using $r_{\mathrm{max}}=300~$pc has negligible effect on the final cluster properties while using $r_{\mathrm{max}}=200~$pc only changes the final properties slightly. Furthermore, it is shown in the top panel of Figure~\ref{fig:sbp_vdp_ndp} that the model SBP deviates from the observed SBP at the edge of the cluster where $r \gtrsim 1500~$arcsec. Attempting to decrease this discrepancy, we tried a simulation with $r_{gc} = 5.5~$kpc (pericenter of the observed 47~Tuc orbit instead of the apocenter at $7.4~$kpc; \citealp{Baumgardt+2019}) and initial tidal radius of $149~$pc. Although this did lower the outermost portion of the model SBP, the discrepancy was still present. This could be because our models do not take into account rotation (see Section~\ref{subsec:caveats}) or because the observations at the edge of the cluster are incomplete due to contamination from field stars (the fit is better for the NDPs in the bottom panel where the observations at the edge of the cluster are taken with Gaia; \citealp{deboer+2019}).

The age of 47~Tuc can be estimated by modeling the properties of the cluster's white dwarf (WD) cooling sequence, comparing its CMD to theoretical isochrones, or fitting stellar evolution tracks to observed eclipsing binaries in the cluster. The estimated ages are uncertain and span a wide range from $\sim 9-14~$Gyr \citep[e.g.,][and references therein]{Dotter+2010,Hansen+2013,VandenBerg+2013,Brogaard+2017,Thompson+2010,Thompson+2020}. Our models predict that the age of 47~Tuc is between $\sim 9~$Gyr to $\sim 12~$Gyr, consistent with these estimates.

\begin{figure}[h]
\includegraphics[width=\columnwidth]{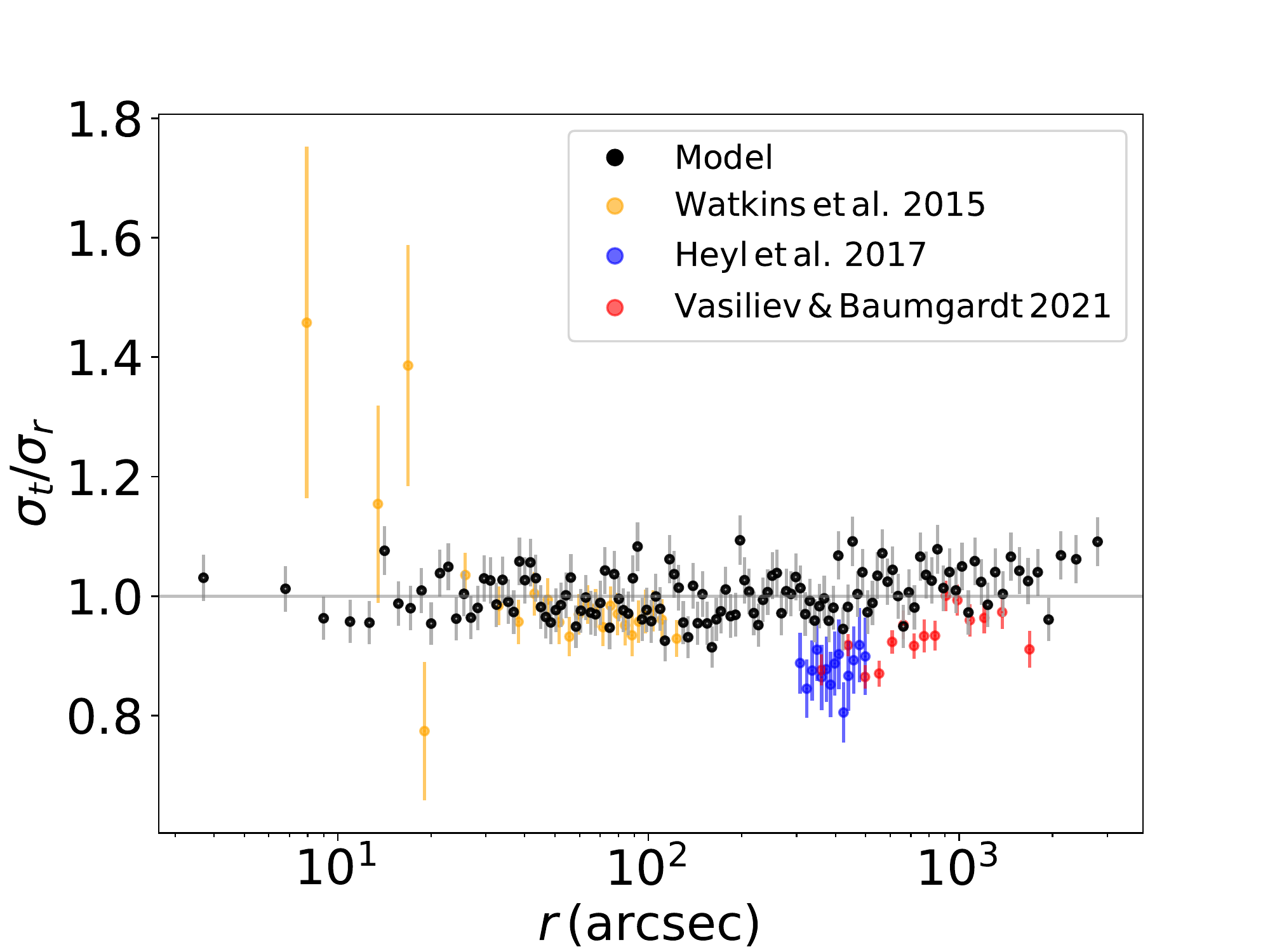}
\caption{Ratio between the tangential and the radial velocity dispersions. Similar to Figure~\ref{fig:sbp_vdp_ndp}, black markers show the best-fit model. Observations \citep{Watkins+2015,Heyl_2017,Vasiliev+2021} show slight tangential velocity anisotropy in the inner part and radial velocity anisotropy in the outer part. The model is isotropic in general but still within the uncertainties of the observational points. \label{fig:sigmatr}}
\end{figure}

Figure~\ref{fig:sigmatr} compares the best-fit model's velocity anisotropy profile to the observations. Some recent studies \citep[e.g.,][]{Vasiliev+2021} have shown that some GCs exhibit velocity anistropy, which may be caused by rotation. In particular, observations of 47~Tuc \citep{Watkins+2015,Heyl_2017,Vasiliev+2021} using both HST and Gaia show small velocity anisotropy both in the inner and outer parts of the cluster. Our best-fit model is consistent with being isotropic and, overall, is within the velocity anisotropy uncertainties of the observations.

\begin{figure*}
\begin{center}
\includegraphics[width=\textwidth]{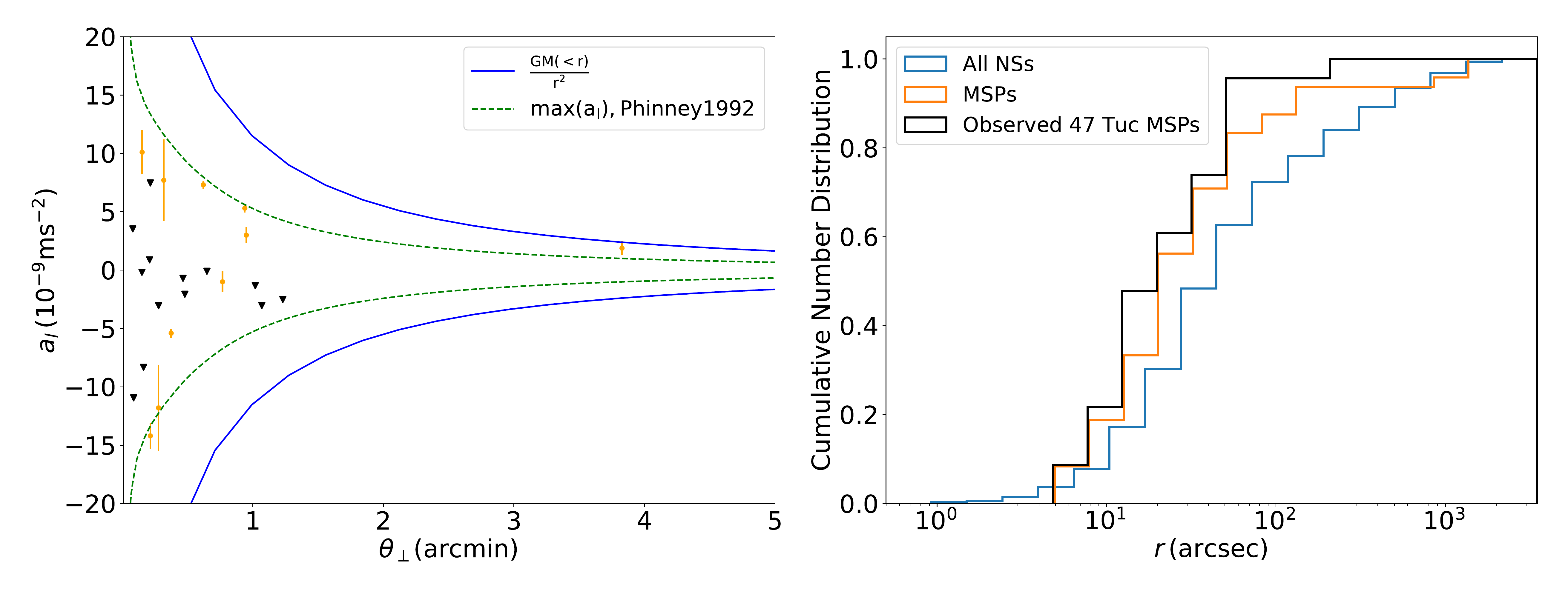}
\caption{Left: Observed pulsar line-of-sight accelerations as a function of angular offsets from the cluster center (orange and black markers; \citealp{Ridolfi+2016,Freire_47tuc_2017,Freire_Ridolfi_2018,Ridolfi+2021}). The black inverted triangles show the acceleration upper limits of isolated pulsars or pulsars in binaries without orbital period data. The orange markers show the accelerations and their uncertainties for pulsar binaries with orbital period timing. The blue solid curves show the maximum possible accelerations from the model cluster potential, and the green dashed curves show the maximum model line-of-sight accelerations. Right: Cumulative number distribution of NSs and MSPs as a function of angular offsets from the cluster center. The black curve is the observed pulsar distribution, the orange curve is the model MSP distribution and the light blue curve is the model distribution for all NSs. \label{fig:psr_acc}}
\end{center}
\end{figure*}

Since 47~Tuc contains the second largest population of MSPs ($28$ observed to date; \citealp{psr_catalog}), 47~Tuc models must be able to explain its pulsar accelerations in the cluster potential. Accurate pulsar line-of-sight accelerations are obtainable from timing analysis of pulsar binary orbital periods. Upper limits on the acceleration are also calculable from the pulsar spin derivatives, in the case of an isolated pulsar or binary with incomplete timing \citep{Ridolfi+2016,Freire_47tuc_2017,Freire_Ridolfi_2018,Ridolfi+2021}. The observed line-of-sight accelerations and upper limits for 47~Tuc MSPs are shown in Figure~\ref{fig:psr_acc} (left panel, orange and black markers\footnote{We adopt the accelerations and their uncertainties for MSPs in binaries from the original timing analysis using slightly different 47~Tuc distance ($4.69~$kpc in \citealp{Freire_47tuc_2017}) as is used in this paper since the distance change has negligible effect on the accelerations \citep[second term on the right side of equation 13 in][]{Freire_47tuc_2017}. We have updated the upper limits for MSPs without orbital period timing using the latest 47~Tuc distance measurement from Gaia \citep{Baumgardt+2021}.}).

\startlongtable
\begin{deluxetable}{C C C C C}
\tabletypesize{\scriptsize}
\tablewidth{3pt}
\tablecaption{Present-Day Properties of the Best-fit Models \label{tab:present_prop}}
\tablehead{
\colhead{} & \colhead{$\mu$} & \colhead{$\sigma$} & \colhead{Obs.} & \colhead{Ref.}
}
\startdata
$\rm{Mass}\,(\rm{M_{\odot}})$ & $1.01\times10^6$ & $1.28\times10^4$ & - & -\\
 $\rm{r_c}\,(\rm{pc})$ & $0.89$ & $0.03$ & - & -\\
 $\rm{r_h}\,(\rm{pc})$ & $6.56$ & $0.15$ & - & -\\
 $\rm{r_{cobs}}\,(\rm{arcmin})$ & $0.35$ & $0.02$ & $0.36$ & (1)\\
 $\rm{r_{hlobs}}\,(\rm{arcmin})$ & $2.71$ & $0.14$ & $3.17$ & (1)\\
 $\rm{N_{BH}}$ & $186$ & $14$ & $1$ & (3)\\
 $\rm{N_{BH-BH}}$ & $3$ & $1$ & - & -\\
 $\rm{N_{NS}}$ & $1368$ & $1$ & - & -\\
 $\rm{N_{LMXB}}$ & $14$ & $2$ & $6$ & (2)(3)(4)\\
 $\rm{N_{MSP}}$ & $54$ & $3$ & $28$ & (5)\\
 $\rm{N_{young\,psr}}$ & $0$ & $0$ & $0$ & (5)\\
 $\rm{N_{CV}}$ & $260(334)$ & $19(21)$ & $43-370$ & (4)(6)\\
 $\rm{N_{BSS}}$ & $100(101)$ & $4(5)$ & $114$ & (7)\\
\enddata
\tablecomments{From top to bottom: Total mass, theoretical core radius, theoretical half-mass radius, observational core radius, observational half-light radius, number of BHs, number of binary BHs, number of NSs, number of LMXBs, number of MSPs (including tidal capture NS--MS binaries), number of young pulsars (spin period $> 30~$ms), number of CVs within theoretical half-mass radius (total number in the model), and number of BSSs within theoretical half-mass radius (total number in the model). The mean values $\mu$ are averages over $3~$Gyr from $\sim9~$Gyr to $\sim12~$Gyr calculated using at least $19$ snapshots, and $\sigma$ are the standard deviations. (1) \citet[][2010 edition]{Harris_1996}, (2) \citet{Heinke+2005}, (3) \citet{MillerJones_2015}, (4) \citet{Bhattacharya+2017}, (5) \citet{psr_catalog}, (6) \citet{RSandoval_2018}, (7) \citet{Parada+2016}. The theoretical core radius is mass-density-weighted and calculated using the method in \citet{Casertano_Hut_1985}. The model observational half-light radius is defined to be the 2D-projected radius that contains half of the cluster's total light, and the model observational core radius is estimated using the method described in \citet{Morscher+2015,Chatterjee+2017}.}.
\end{deluxetable}

We estimate the maximum line-of-sight accelerations of model pulsars using equation 2.5 in \citet{Phinney_1992},

\begin{equation}
    \max |a_l| \approx 1.1\frac{GM_{cyl}(<R_{\perp})}{\pi R_{\perp}^2},
\end{equation}
where $M_{cyl}$ is the mass of the cluster within a cylindrical tube of projected radius $R_{\perp}$ from the cluster center along the line-of-sight. This is shown by the green curves in Figure~\ref{fig:psr_acc} (left panel). We also include the distribution of maximum possible acceleration allowed by the cluster potential in Figure~\ref{fig:psr_acc} (left panel, blue curves), calculated by $\frac{GM(<r)}{r^2}$, where $M$ is the mass within a sphere of radius $r$. To $1\,\sigma$ uncertainty, all but one of the observed pulsars have line-of-sight accelerations within the bounds allowed by the model potential, demonstrating further that our best-fit model is consistent with the observations. The only outlier at $\sim 4~$arcsec is still enclosed by the maximum possible \emph{total} acceleration allowed by the cluster potential.

The right panel of Figure~\ref{fig:psr_acc} shows the cumulative number of all model NSs (light blue curve), model MSPs (orange curve), and observed 47~Tuc MSPs (black curve) as a function of the projected distance from the cluster center. All of the observed 47~Tuc pulsars are MSPs with spin periods $< 30~$ms. As expected, the observed MSP distribution agrees better with the modeled MSP distribution than the total NS distribution. The total NS population is distributed slightly further out because MSPs are formed through extended periods of mass accretion and are slightly more massive ($\sim 0.1~\rm{M_{\odot}}$ more) than most other NSs. Mass segregation leaves the MSPs closer to the cluster center than typical NSs. Also NSs closer to the cluster center are more dynamically active, so they more easily acquire companion stars through encounters and become MSPs \citep{Ye_msp_2019}.

\begin{figure*}
\begin{center}
\includegraphics[width=\textwidth]{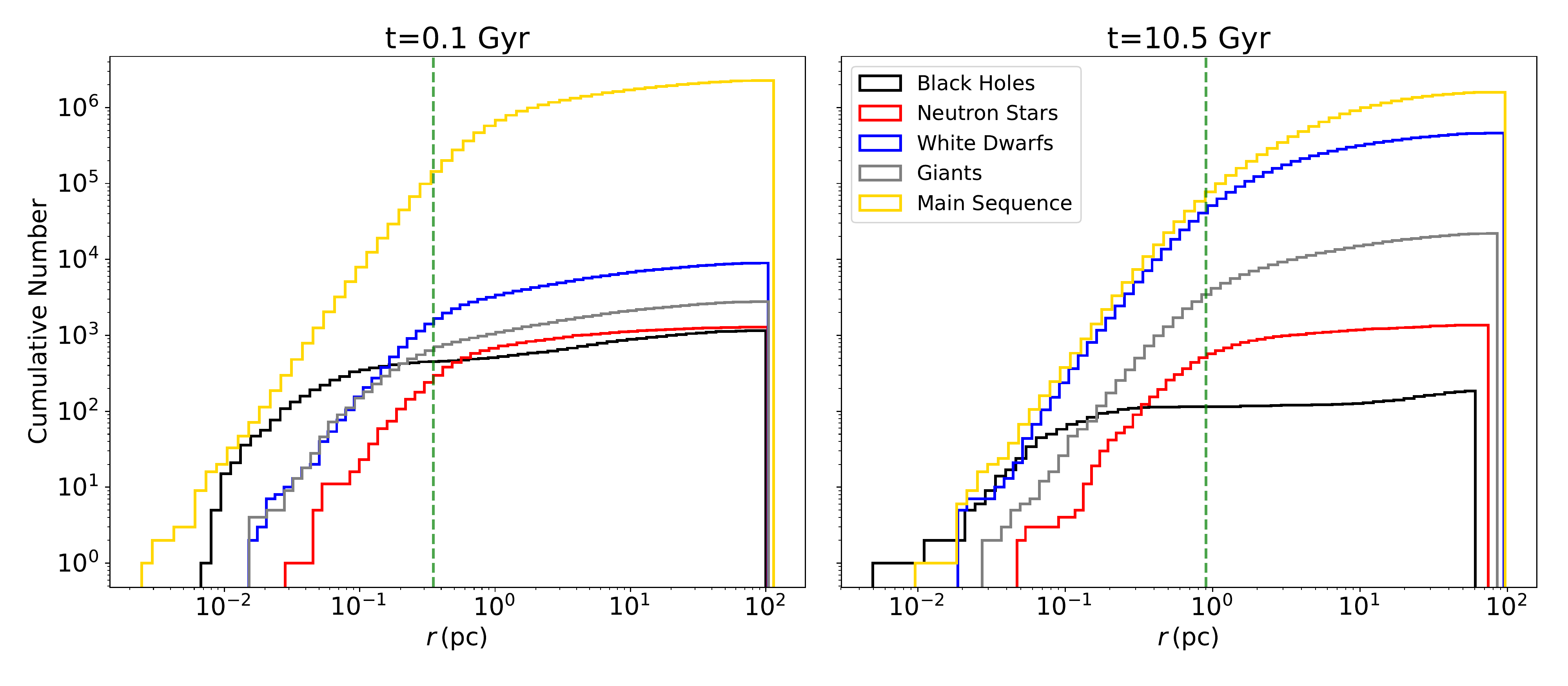}
\caption{Cumulative number distributions of all stellar populations at $0.1~$Gyr (left panel) and at $10.5~$Gyr (right panel) as the function of their distances to the cluster center in three dimension. The vertical dashed lines show the theoretical core radii.\label{fig:star_radius}}
\end{center}
\end{figure*}

Table~\ref{tab:present_prop} lists the properties of the best-fit 47~Tuc models at the present day, including the total numbers of compact objects and blue straggler stars (BSSs). In addition, Table~\ref{tab:general_prop} in Appendix~\ref{app:allsimulations} shows the more general mean and standard deviation values of the same properties for best- and near-fit models from simulations 1-3 in Table~\ref{tab:allsim}. The inclusion of simulation 3 with a high initial binary fraction leads to large standard deviations for the numbers of some compact objects. Figure~\ref{fig:star_radius} shows the cumulative number distributions of all stellar populations (colors denoted in legend) at an early and a late time of the 47~Tuc simulation. As expected, the cluster core is dominated by BHs due to mass segregation, mixing with the most numerous MS stars initially \citep[see also][]{Kremer+2020catalog}, and also WDs at late times. This is also reflected in the cluster's SBP, where it is flat within $\sim 20~$arcsec, consistent with the features of a non-core-collapsed cluster.

\section{Exotic objects in 47~Tuc}\label{sec:compact}
Here we report the number of exotic objects in the best-fit 47~Tuc model, which can be used to estimate the total numbers of compact objects, including BHs, MSPs, LMXBs and CVs, and numbers of BSSs, in 47~Tuc. Most compact object binaries discussed below, including the binary BHs in Section~\ref{subsec:bh_merger}, are formed through dynamical interactions; only $\sim10\%$ of MSPs and $\sim35\%$ of CVs are formed in primordial binaries. This high fraction of dynamical compact object binaries is not surprising given the high density of the 47~Tuc models and the low initial primordial binary fraction we adopted, which is motivated by observations of 47~Tuc (\citealp{Milone+2012}; see also Section~\ref{subsec:bfrac}).

\subsection{Pulsars}\label{subsec:pulsar}
A total of $28$ pulsars (all of them MSPs) have been observed in 47~Tuc and $18$ of them are in binaries \citep{psr_catalog}. Among the MSPs in binaries, $6$ of them are ``black widow" MSPs with low-mass, planet- or white/brown dwarf-like stellar remnant companions ($\lesssim 0.05~\rm{M_{\odot}}$), and $3$ are ``redback" MSPs with MS star-like, non-degenerate companions ($\sim 0.1-1~\rm{M_{\odot}}$; \citealp[][figures and references therein]{Robert_2013,Kaplan+2018,Strader+2019,psr_catalog}). Both black widows and redbacks have tight orbital periods of $\lesssim 1~$day.

\begin{figure}
\begin{center}
\includegraphics[width=\columnwidth]{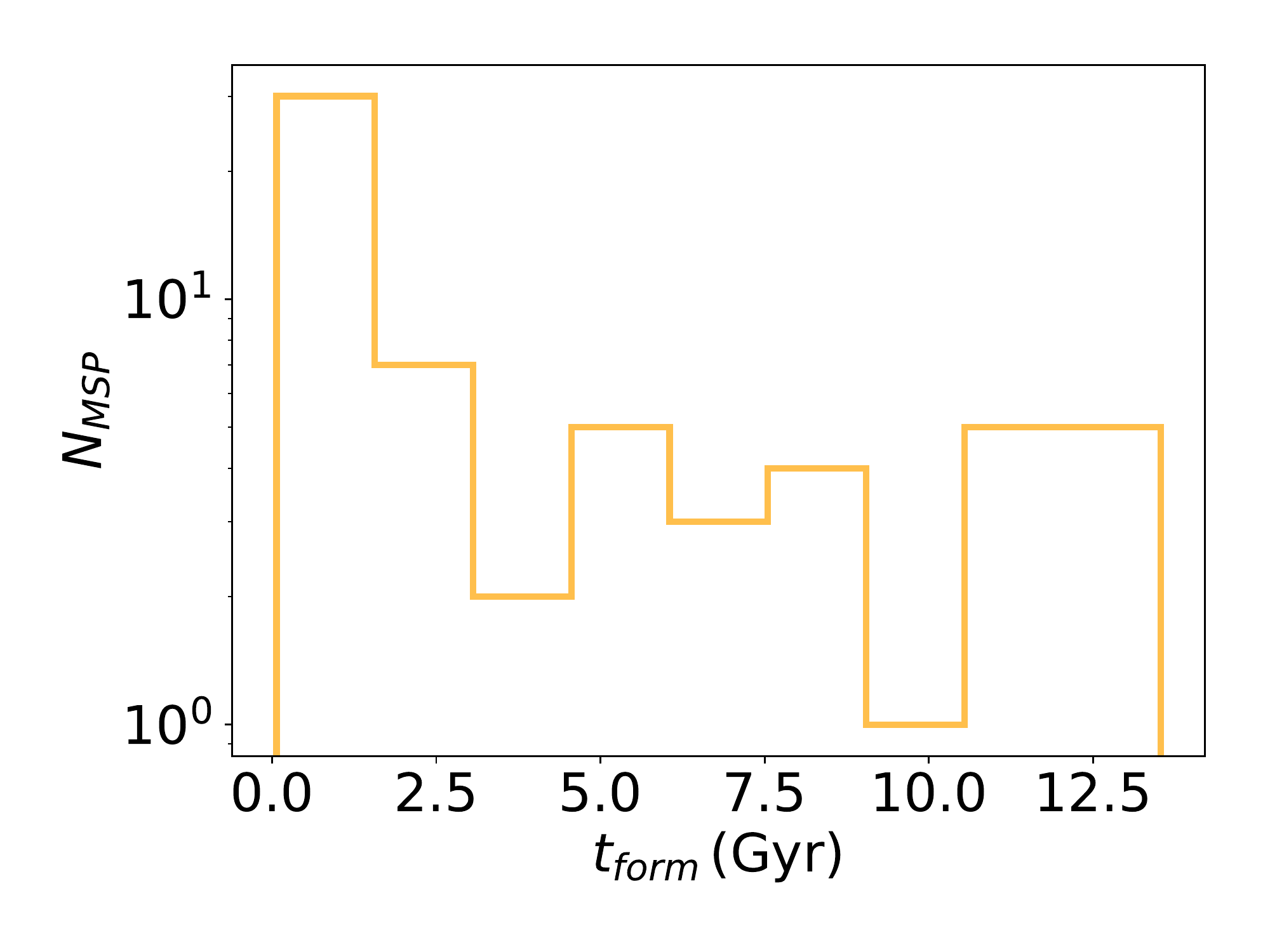}
\caption{Formation time distribution of all MSPs ever formed in the best-fit 47~Tuc simulation in a Hubble time. \label{fig:msp_tform}}
\end{center}
\end{figure}

On average between $\sim9-12$ Gyr, there are a total of $\sim 49$ MSPs per snapshot, $\sim 23$ in binaries and $\sim 26$ isolated\footnote{We count the $26$ MSPs with very low-mass ($\lesssim 0.01~\rm{M_{\odot}}$) CO/ONeMg WD companions, where the WDs are likely unstable as isolated MSPs. This is because there may exist minimum masses for WDs with sufficiently high temperature ($T\gtrsim10^4$K) during mass transfer where there is no equilibrium solution of the WDs' equation of states and the WDs may ``evaporate" \citep{Bildsten_2002,Deloye_Bildsten_2003}.}. Specifically at $10.5~$Gyr, there are $26$ isolated MSPs, $22$ binary MSPs and $0$ young pulsar (spin period $> 30~$ms). Most of the MSPs ($32$ of them) are formed through collisions with giant stars. Most of these giant star collisions occur between a giant star and a MS star, where the giant star's core becomes a NS in subsequent binary and stellar evolution (see also Section~\ref{subsec:giant}). A few of the giant star collisions are between a WD and a giant star. Figure~\ref{fig:msp_tform} shows the distribution of formation times of all MSPs formed in a Hubble time in the best-fit 47~Tuc simulation. At early times of the cluster's evolution ($\gtrsim 3$ Gyr), collisions with giant stars play a significant role ($\sim 80\%$) in forming the binaries which host MSPs later, or lead to more binary-mediated dynamical encounters and subsequent MSP formation. This is shown by the peak in the formation time distribution in Figure~\ref{fig:msp_tform}. After $\sim 3$ Gyr, more than half of the BHs have been ejected and the cluster is denser than earlier. The higher density leads to enhanced dynamical interactions and the formation of subsequent MSP-host binaries through exchange encounters or tidal capture. In total, $\sim 40\%$ of MSPs formed after 3 Gyr.

\begin{figure}
\begin{center}
\includegraphics[width=\columnwidth]{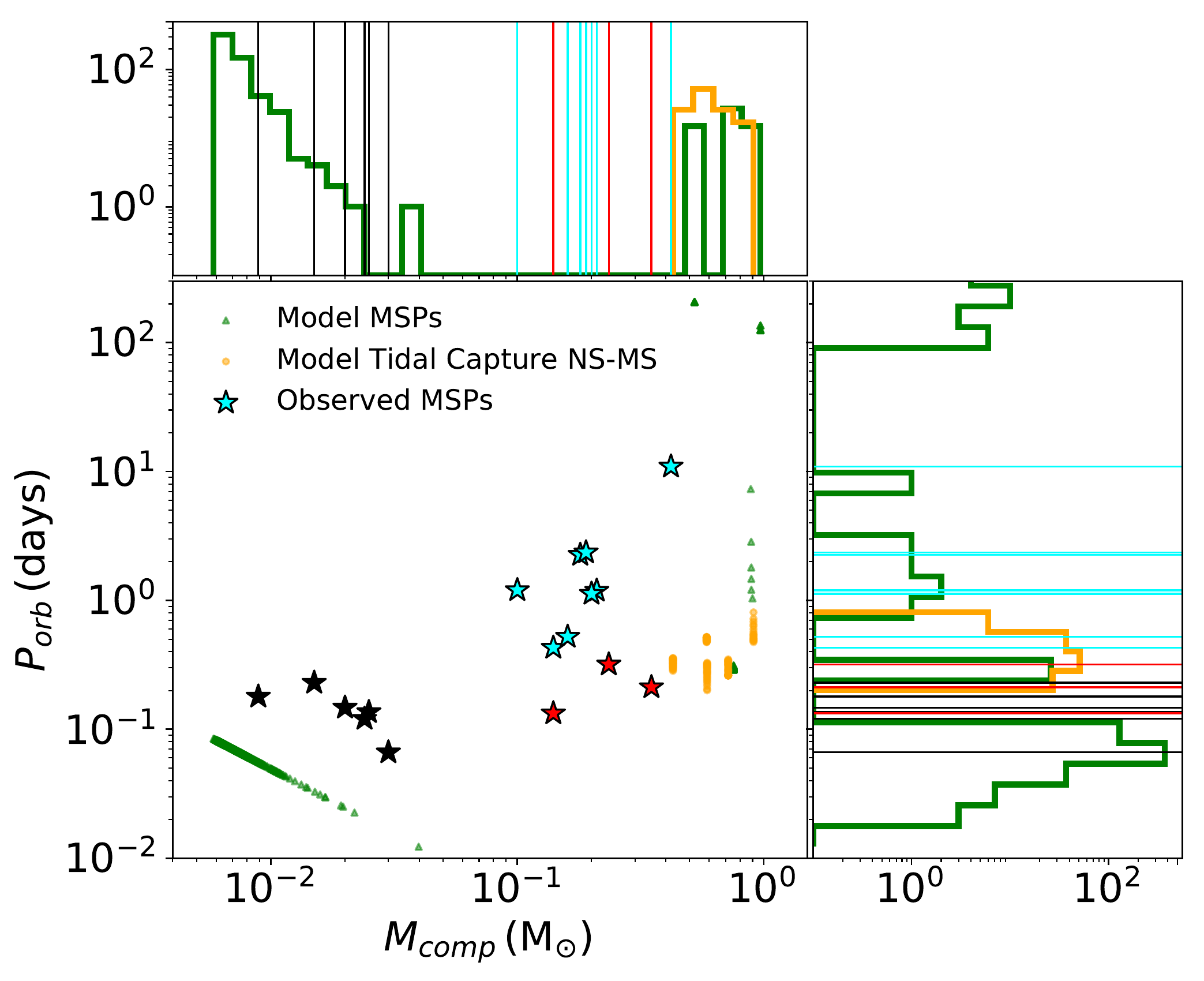}
\caption{Orbital periods versus companion masses of MSP binaries in 47~Tuc. The stars show the observed 47~Tuc MSPs. Black stars are black widows, red stars are redbacks and blue stars are ``normal" MSPs with WD companions. Model MSPs are shown by green triangles and NS--MS star binaries from tidal capture are shown by orange circles. \label{fig:psr_pbm}}
\end{center}
\end{figure}

In Figure~\ref{fig:psr_pbm}, we further compare the orbital periods and companion masses of \textit{all} model MSP binaries and tidal capture NS--MS star binaries from $9-12~$Gyr to the observations. Roughly speaking, \texttt{CMC} models can produce all three groups of pulsar binaries that bracket the ones observed (black widows, redbacks and MSPs with WD companions, respectively). It is important to point out that tidal capture can produce redback-like NS--MS star binaries (orange circles in Figure~\ref{fig:psr_pbm}, $~5$ per snapshot) with similar companion masses and orbital periods, but these $5$ NSs are not spun up to MSPs in the models. In a realistic scenario during tidal capture, a MS star is likely to expand from the injection of orbital energy and fill its Roche Lobe \citep[e.g.,][]{Kochanek_1992}. Thus it is reasonable to assume that a captured, puffed-up MS star can transfer mass to a NS and spin it up to a MSP. However, \texttt{CMC} currently lacks detailed treatments of MS stars in tidal capture interactions, and \texttt{COSMIC} does not include the latest pulsar physics such as pulsar irradiation \citep[e.g.,][and references therein]{Chen+2013,Ginzburg_Quataert_2020}. We will consider these details and study model MSPs and tidal-captured NS--MS star binaries that overlap with the observed pulsar binaries in future works.

In total, the best-fit model produced $54$ MSPs including the NS--MS star binaries formed in tidal capture interactions. This number is larger than the actual observed number of MSPs in 47~Tuc likely because pulsar searches are limited by selection effects such as the interstellar dispersion and scattering, and survey thresholds \citep[e.g.,][and references therein]{Lorimer_2008}. Our estimate is consistent with the upper-limit estimates of the potential total number of MSPs in 47~Tuc from previous studies \citep{Heinke+2005,Abdo_2009}.

\newpage
\subsection{Low-mass X-ray Binaries and Cataclysmic Variables}\label{subsec:LMXB_CV}
Observations have detected $370$ X-ray sources within $2'79$ of 47~Tuc using the Chandra Observatory \citep{Heinke+2005,Bhattacharya+2017} and found $\sim 5$ quiescent LMXBs \citep{Heinke+2005}. Later studies have further confirmed $43$ CVs \citep{RSandoval_2018} and found $1$ candidate BH LMXB \citep{MillerJones_2015,Bahramian+2017}.

We define LMXBs as BHs or NSs in mass-transferring binaries with WD, giant star, or MS star companions in our models. There are $\sim 14$ LMXBs on average per snapshot in the time scale of $3~$Gyr in our best-fit models, where $\sim 3$ are NS--MS star binaries, $\sim 5$ are BH-WD binaries, $\sim 1$ is a BH-giant star binary, and $\sim 5$ are BH-MS star binaries. There are a similar number of LMXBs at $10.5~$Gyr. Similar to LMXBs, we define CVs as mass-transferring WD binaries with MS star companions \citep[e.g.,][]{Kremer+2021WD}. There are on average $\sim 334$ CVs at $\sim 9-12~$Gyr, and a similar number at $10.5~$Gyr. Most of the CVs ($\sim 260$) and almost all of the LMXBs identified in our models are within the theoretical half-mass radius, consistent with the detected number of X-ray sources. These are only general comparisons between the number of LMXBs and CVs in models and from observations, since we do not consider properties such as duty cycles and luminosities of these mass-transferring binaries. It is not surprising that our models predict more LMXBs and CVs than are observed. The observations of LMXBs and CVs are also limited by selection effects, including the stellar crowdedness in GCs and the faintness of some of these binary systems (\citealp{Heinke_2003} estimated that there are $7$ times more quiescent LMXBs than bright LMXBs in GCs). The number of CVs from the best-fit 47~Tuc models is consistent with the estimates in \citet{Ivanova_2006} and \citet{RSandoval_2018}, which showed that clusters similar in mass to 47~Tuc can have $\sim 300$ CVs. This provides more support for the number estimates of compact objects in our 47~Tuc model.

\subsection{Black Holes}\label{subsec:BH}
47~Tuc has one observed BH X-ray binary candidate \citep[47 Tuc X9;][]{MillerJones_2015, Bahramian+2017}. Recently, \citet{Weatherford+2020} used mass segregation measurements to infer a total population of $43^{+146}_{-37}$ ($2\sigma$ uncertainties) stellar-mass BHs in 47 Tuc, a few percent of which may be found in mass-transferring binaries similar to the X9 source \citep[e.g.,][]{Kremer+2018}. It has also been suggested that 47 Tuc may host an intermediate-mass BH \citep{Kiziltan_2017} with mass of $2300_{-850}^{1500}~\rm{M_{\odot}}$.

\begin{figure}[h]
\begin{center}
\includegraphics[width=\columnwidth]{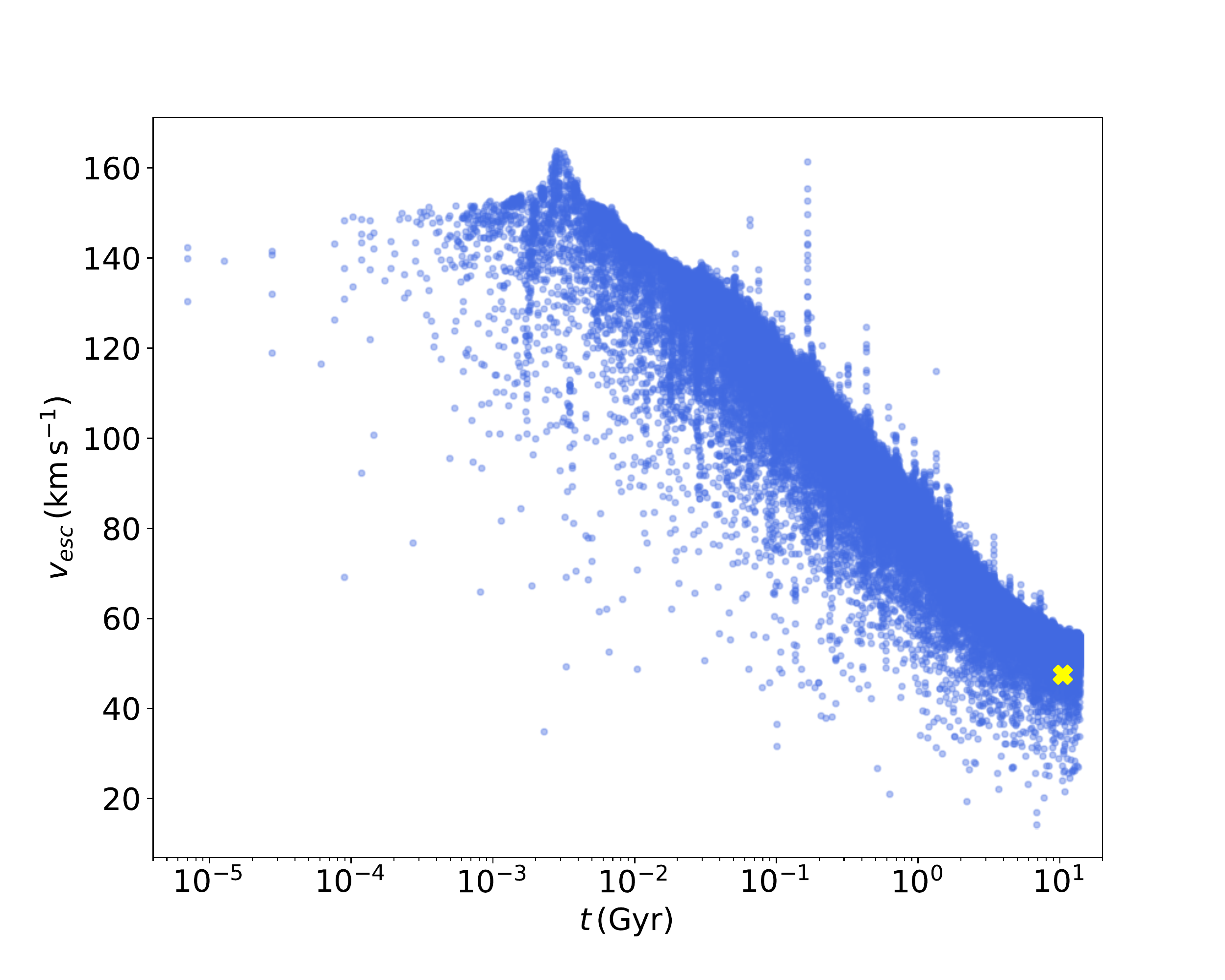}
\caption{Escape velocities during strong interactions (single-single, binary-single and binary-binary encounters) as a function of time. The escape velocities of the cluster decrease about $3$~times from $\sim 160~\rm{km\,s^{-1}}$ to $\sim 60~\rm{km\,s^{-1}}$ over a Hubble time. The yellow cross marker shows the central escape velocity ($47.5~\rm{km\,s^{-1}}$) of 47~Tuc from Holger Baumgardt's catalog (\url{https://people.smp.uq.edu.au/HolgerBaumgardt/globular/parameter.html}), assuming $10.5~$Gyr. \label{fig:v_esc}}
\end{center}
\end{figure}

Initially, the model cluster produced $\sim 1270$ BHs. Most of them are ejected later through dynamical interactions, as the escape velocities of the cluster decrease significantly over a Hubble time (Figure~\ref{fig:v_esc}). On average there are $\sim 186$ BHs and $\sim 3$ binary BHs over a $3~$Gyr timescale from $\sim9~$Gyr to $\sim12~$Gyr, while there are $182$ BHs and $2$ binary BHs at the best-fit model (Table~\ref{tab:present_prop}). The number of retained BHs at present is consistent with the $2\sigma$ upper limit, and the total mass of the BHs per snapshot ($\sim 2350~\rm{M_{\odot}}$) is consistent with the $1\sigma$ upper limit in \citet{Weatherford+2020}. However, the BH mass in our best-fit models is twice as large as in \citet{Henault-Brunet+2019}, indicating different IMFs may vary the final retained BH mass significantly (although they did not specify the maximum and minimum masses used in their IMF). Furthermore, the total mass of BHs in the 47~Tuc model is about the same as the mass of a potential intermediate-mass BH in the cluster \citep{Kiziltan_2017}. However, since the pulsar accelerations in 47~Tuc can be reproduced with a group of stellar-mass BHs, the existence of an intermediate-mass BH is not necessary \citep[also see][]{Mann+2019}. We will study how an intermediate-mass BH affects the properties of a cluster in future works.

\subsection{Blue Stragglers}\label{subsec:BSS}
We define model BSSs based on the MS star turn-off (TO) luminosity, where the model MS stars reach the highest temperature. Temperatures of isolated stars are calculated using the stars' luminosities and radii (\texttt{COSMIC} outputs) assuming a black body radiation model, and temperatures of binary stars are calculated from the combined luminosity-weighted temperature of the component stars, where the total luminosity of a binary is the sum of the luminosities of its component stars \citep[][equation 1]{Weatherford_2018}. To avoid contamination from normal MS stars, we specify BSSs as upper-MS stars with temperature $T > T_{TO}$ and luminosity $L > 2L_{TO}$, either isolated or in binaries \citep[e.g.,][]{Kremer+2020catalog}. 

\begin{figure}
\begin{center}
\includegraphics[width=\columnwidth]{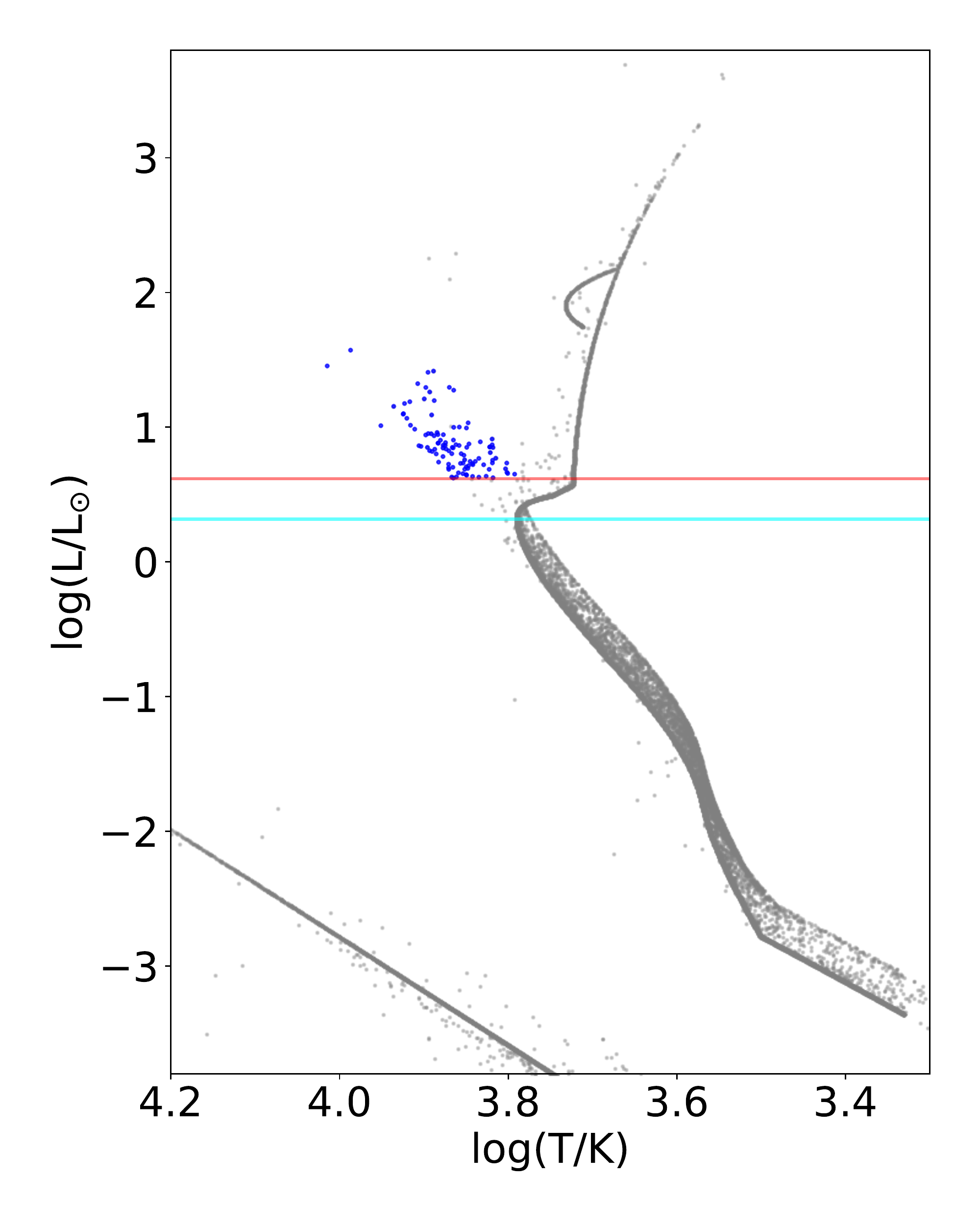}
\caption{H-R diagram of the best-fit model. Blue dots show the BSSs. There are $108$ BSSs at this time. The aqua line shows the turn-off luminosity $L_{TO}$ of the MS stars, and the red line is at twice the turn-off luminosity. The BSSs are selected from upper-MS stars with $L > 2L_{TO}$ and $T > T_{TO}$. \label{fig:bss}}
\end{center}
\end{figure}

\citet{Parada+2016} detected $114$ BSSs using UV data taken from HST of 47~Tuc at $\lesssim 160~$arcsec. In our models, we find a total of $\sim 101$ BSSs per unit time, and $\sim 100$ of them are within the theoretical half-mass radius (see Figure~\ref{fig:bss} and Table~\ref{tab:present_prop}), less than the observed number of BSSs. This may be because \texttt{COSMIC} under-produces BSSs through the mass transfer channel \citep{Leiner_Geller_2021}. Most BSSs ($\gtrsim 95\%$) in the 47~Tuc models are formed in direct collisions of MS stars. In addition, the $2\sigma$ upper limit of the number of BSSs (Table~\ref{tab:present_prop}) is very close to the observed number of BSSs, suggesting that the potential under-production of BSSs through the mass transfer channel is only a small effect in 47~Tuc.

\begin{figure}
\begin{center}
\includegraphics[width=\columnwidth]{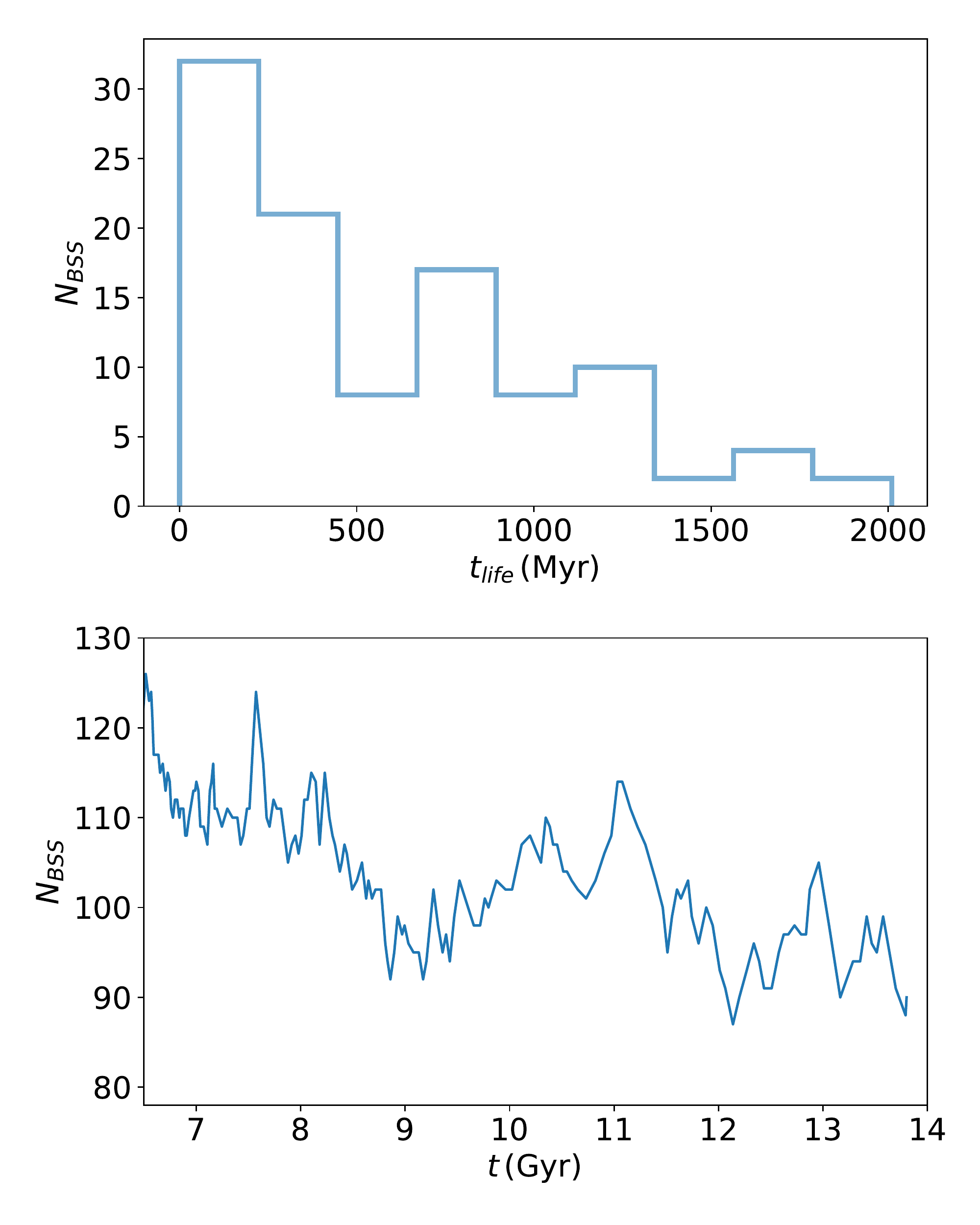}
\caption{Upper panel: Lifetime distribution of BSSs in the best-fit model as shown in Figure~\ref{fig:bss}. All the BSSs have a lifetime $\lesssim 2$ Gyr. Bottom panel: Number of BSSs as a function of time at late times of the cluster's evolution. \label{fig:nbss_lifetime}}
\end{center}
\end{figure}

The lifetime of BSSs in the best-fit model is also shown in Figure~\ref{fig:nbss_lifetime} (upper panel). All of them have lifetimes $\lesssim 2$ Gyr. With the short lifetimes and the continuing formation of BSSs through stellar collisions, the average number of BSSs at late times of the cluster's evolution (Figure~\ref{fig:nbss_lifetime}; bottom panel) is quite stable.

\section{Discussion} \label{sec:discuss}
 \begin{figure}
\begin{center}
\includegraphics[width=\columnwidth]{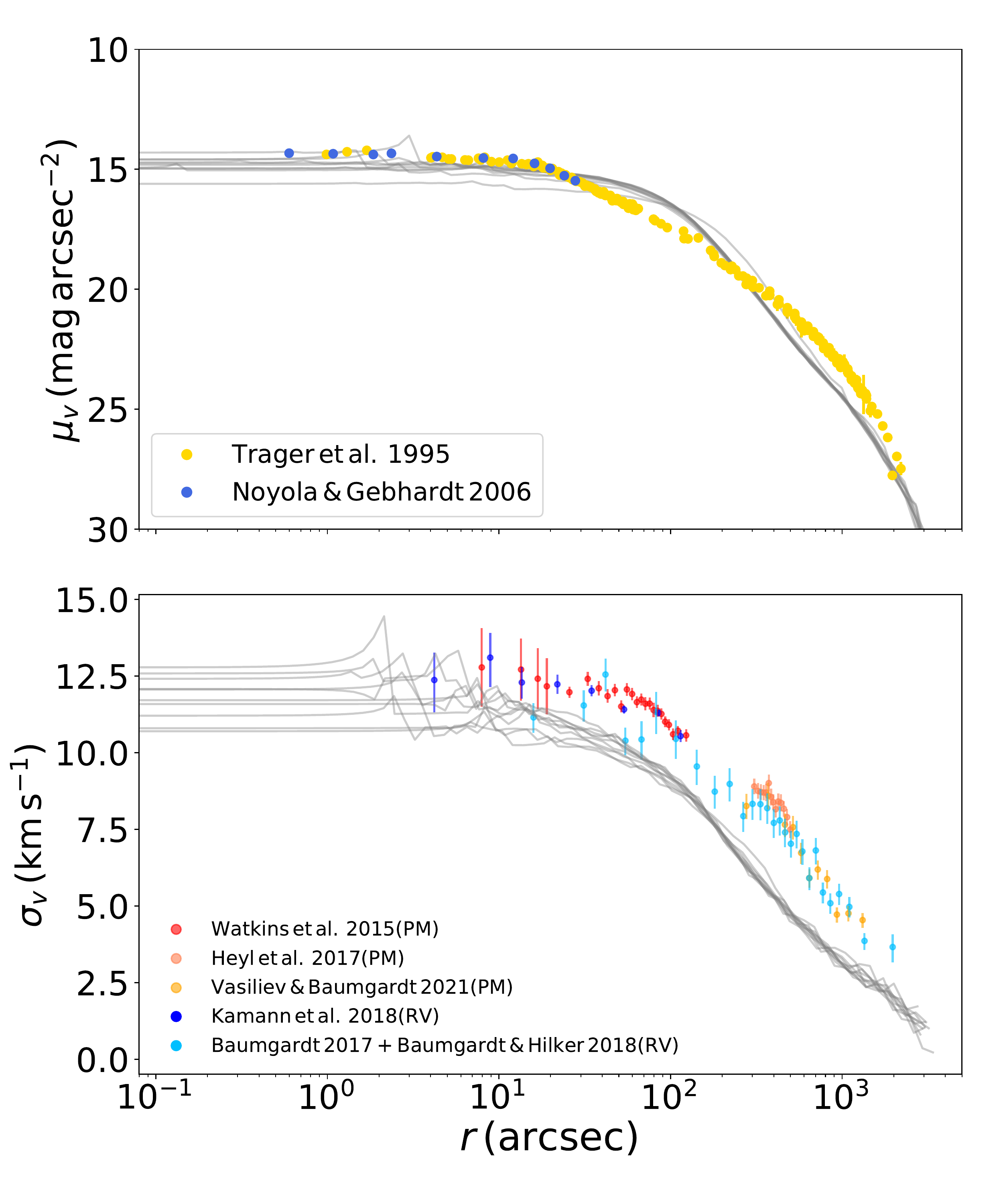}
\caption{Top $10$ closest-fit models for 47~Tuc (gray curves) from the \texttt{CMC} Cluster Catalog \citep{Kremer+2020catalog}. Nine of them are different time steps from the same simulation, where initially it has a canonical mass function from \citet{Kroupa2001}, number of stars $N = 1.6\times10^6$, King concentration parameter $W_0 = 5$, virial radius $r_v = 1~$pc, and binary fraction $f_b = 5\%$. The metallicity is $Z = 0.002$ and the Galacticocentric distance is $8~$kpc. The last model is from a simulation with initial $r_v = 2~$pc and the other initial conditions the same as the above simulation. \label{fig:maingrid_massive}}
\end{center}
\end{figure}

\subsection{King Profile versus Elson Profile}\label{subsec:profile}
Many previous studies have shown that a King profile \citep{King1966} may not be the best match for the observed SBPs of the majority of GCs in the Milky Way because of its sharp cutoff of the density distribution close to the tidal radius \citep[e.g.,][]{McLaughlin+2005sbp,Miocchi+2013,deboer+2019,Henault-Brunet+2019}. Instead, the Wilson profile \citep{Wilson_1975}, which is more spatially extended than the King profile, is shown to fit equally well or better where the observed SBPs exhibit a smoother decrease in the outermost radial density \citep[e.g.,][]{McLaughlin+2005sbp,Miocchi+2013,deboer+2019}. More recently, studies have demonstrated that multi-mass models from the \texttt{LIMEPY} family \citep{Gieles_Zocchi_2015} represent some GCs better than the King or the Wilson profile, including 47~Tuc \citep{deboer+2019,Henault-Brunet+2019}. \texttt{LIMEPY} models are isothermal in the central regions and are described by polytropes in the outer regions. They can reproduce King or Wilson profiles with certain choices of truncation parameter. Furthermore, it has been shown that young star clusters in the Large Magellanic Cloud are better described by an Elson profile with no apparent tidal truncation \citep{Elson1987,Mackey_Gilmore_2003}. The Elson profile is a generalization of the Plummer profile \citep{Plummer_1911} that arises when $\gamma = 4$ in equation~(8). The Elson profile is similar to the King profile in the inner regions of a cluster, but exhibits larger density or surface brightness at radii approaching the tidal radius (see Figure~4 in \citealp{Mackey_Gilmore_2003}). A typical range of $\gamma$ for young star clusters in the Large Magellanic Cloud is $\sim 2.2-3.2$ \citep{Mackey_Gilmore_2003,PZwart+2010}, close to what we adopt for 47~Tuc.

\begin{figure}
\begin{center}
\includegraphics[width=\columnwidth]{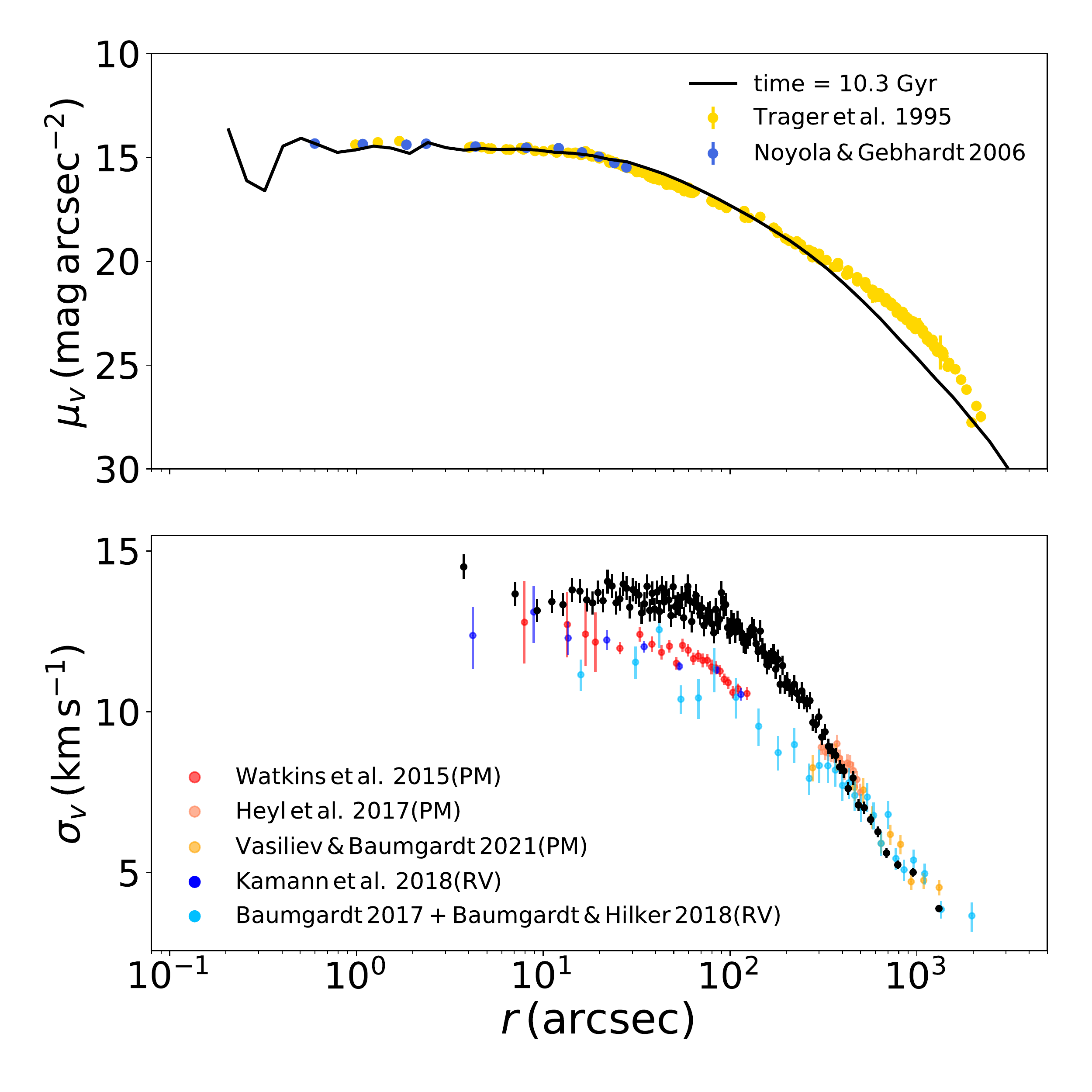}
\caption{A massive \texttt{CMC} model with a King profile (the black curve and black markers). This model has initial number of stars $N = 2\times10^6$, initial King concentration parameter $W_0 = 7.5$, and initial virial radius $r_v = 1.4~$pc. Its initial binary fraction, IMF, metallicity and Galactocentric distance are the same as the best-fit 47~Tuc models (Table~\ref{tab:IC}). These initial conditions are chosen such that the model SBP can match the observed 47~Tuc SBP. \label{fig:king_sbp}}
\end{center}
\end{figure}

Figure~\ref{fig:king_sbp} shows how 47~Tuc's observed SBP and VDP are fit by a massive model with a King profile (as opposed to the best-fit model using an Elson profile shown in Figure~\ref{fig:sbp_vdp_ndp}). The top panels of Figure~\ref{fig:sbp_vdp_ndp} and \ref{fig:king_sbp} show that models using both profiles can fit the observed SBP. However, the King profile model starts to deviate from observations at around $300~$arcsec, which encloses $\sim 20\%$ of the model's total mass. We also explored whether varying the tidal radius of the King models could improve their fit to the outer part of the SBP. Changing the Galactocentric distance from 47~Tuc's apocenter $7.4~$kpc to its pericenter $5.5~$kpc \citep{Baumgardt+2019}, thereby shrinking the tidal radius of the cluster, does not affect the SBP. This indicates that models with a King profile are underfilling. Furthermore, we also ran a set of small models (initial masses $\sim 2\times10^5~\rm{M_{\odot}}$) with different initial King concentration parameters $W_0= 3, 5, 7, 9$ in order to alter the cluster potential's shape. We find that unless $W_0$ is very large ($\gg9$), there is no apparent change in the final SBPs. More importantly, the model VDP does not match the observed VDP of 47~Tuc (Figure~\ref{fig:king_sbp}, bottom panel). Figure~\ref{fig:maingrid_massive} further demonstrates that a flatter density profile than the King profile is needed to explain the observed VDP at radii $\gtrsim 50~$arcsec. Taking all these into account, we adopt the Elson profile for better fits to the outer part of the cluster in this study.

\subsection{Initial Mass Function}\label{subsec:imf}
Whether or not the IMF is universal for GCs is still an ongoing debate. For example, \citet{Marks+2012} found that stellar IMFs are more top-heavy (more high-mass stars) as a cluster's metallicity becomes lower and the gas density of the pre-GC cloud becomes larger. Later, \citet{Haghi+2017} showed that top-heavy IMFs can better explain the relation between the observed mass-to-light ratios and the metallicities of GCs in the Andromeda galaxy. However, in contrast to the suggested initial power-law slope for the mass function of massive stars ($\gtrsim 1~\rm{M_{\odot}}$) in 47~Tuc \citep{Marks+2012}, both \citet{Giersz47Tuc2011} and \citet{Henault-Brunet+2019} found that steeper power-law slopes (comparing to \citealp{Kroupa2001}) better match the 47~Tuc observations. The power-law slope we adopt for high-mass stars ($\alpha_2 = 2.8$ for $0.8 \leq m/\rm{M_{\odot}}$ in Table~\ref{tab:IC}) is roughly within the $95\%$ uncertainties of the Kroupa IMF ($\alpha_2 = 2.3\pm0.3$ for $0.5 \leq m/\rm{M_{\odot}} < 1.0$ and $\alpha_3 = 2.3\pm0.7$ for $1 \leq m/\rm{M_{\odot}}$; \citealp{Kroupa2001}) except that we use a different break mass. This indicates that 47~Tuc may be top-light.

\begin{figure}
\begin{center}
\includegraphics[width=\columnwidth]{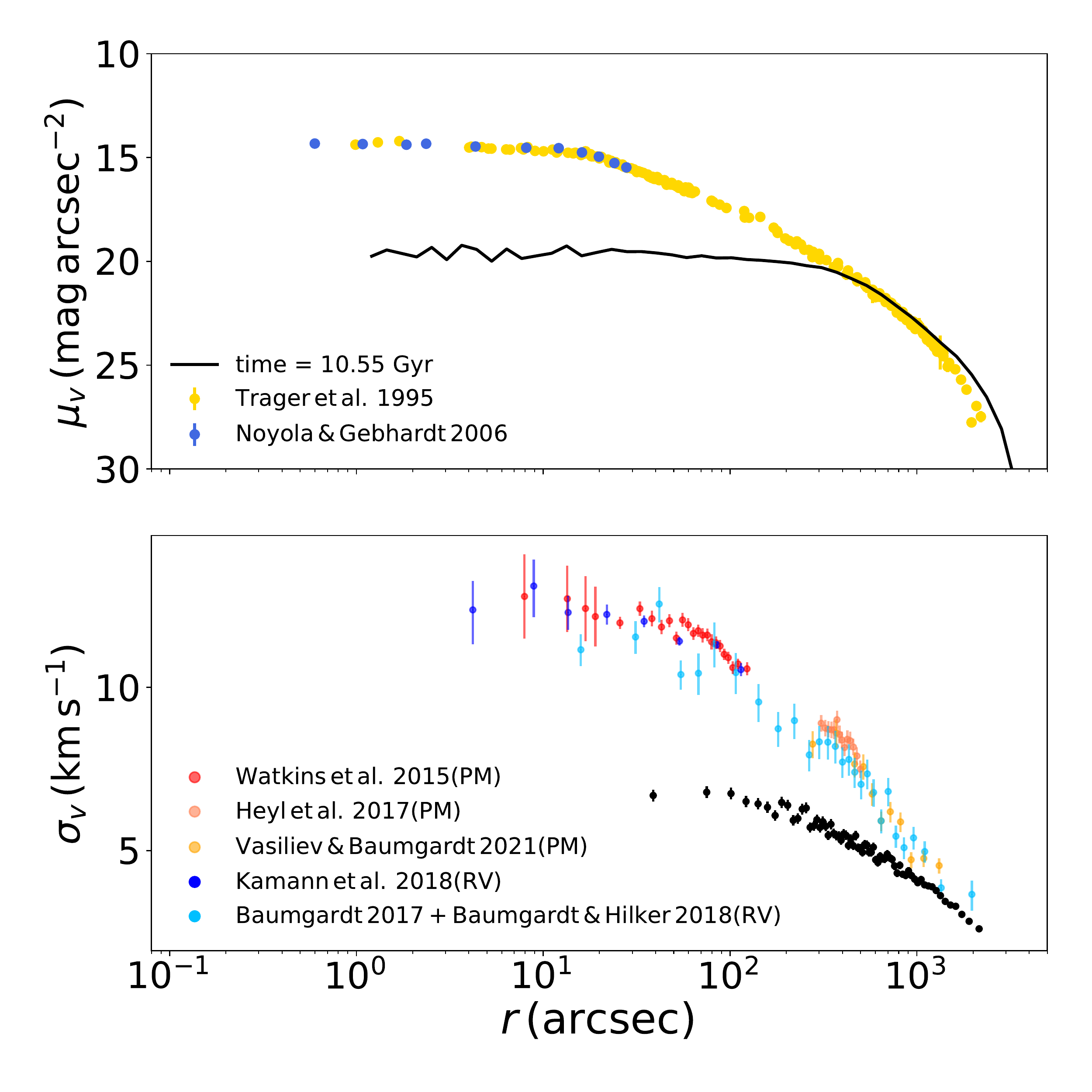}
\caption{A massive model at $t = 10.55~$Gyr with initial $\alpha_2 = 2.3$ and initial $N = 1.96e6$ (the black curve and black markers). The initial $N$ is lowered to keep the initial mass of this model the same as that of the best-fit model. All other initial conditions are the same as the best-fit model. \label{fig:imf2.3}}
\end{center}
\end{figure}

We also ran simulations with $\alpha_2 = 2.3, 1.8$ (the initial mass and other initial conditions are the same as in Table~\ref{tab:IC}) to further check if a power-law slope for massive stars more consistent with the Kroupa IMF or a top-heavy IMF is allowed. Figure~\ref{fig:imf2.3} shows the SBP and VDP of a model at $10.55~$Gyr with an initial $\alpha_2 = 2.3$. Because the shallower IMF increases the number of massive stars and BHs, the model is puffier (having larger core radius from BH burning; \citealp[e.g.,][]{Kremer+2019size}) and does not fit either the observed SBP or VDP. The simulation with initial $\alpha_2 = 1.8$ dissolved before $5~$Gyr due to even more extreme BH burning \citep[e.g.,][]{Chatterjee_2010,Weatherford+2021}.

It is significant to point out that a top-light IMF may not be the only way to produce a good fit for 47~Tuc. There may be a degeneracy between having a top-light IMF or instead, having a smaller initial virial radius $r_v$, both can reduce the number of BHs and the cluster core radius. A smaller initial $r_v$ corresponds to a shorter two-body relaxation time \citep{Spitzer_1987} and a faster dynamical clock for the cluster \citep{Kremer+2019size}. Thus a model with smaller $r_v$ will process and eject its BHs faster through dynamical encounters in the same time (e.g., $13~$Gyr). However, for 47~Tuc and similar simulations with large initial masses ($\sim 2.5\times10^6~\rm{M_{\odot}}$), a simulation with smaller initial $r_v$ of $\lesssim 1~$pc takes a much longer time than simulations with $r_v > 1~$pc ($> 2~$months compared to $2~$weeks). Therefore we do not explore small $r_v$ models in this paper, and leave them for future studies \citep[e.g.,][]{Weatherfordinprep}.

Furthermore, \citet{Henault-Brunet+2019} found that a power-law with $\alpha_1 = 0.52$ for $m/\rm{M_{\odot}} < 0.5$ and $\alpha_2 = 1.35$ for $0.5 < m/\rm{M_{\odot}} < 1$ matched their 47~Tuc models. This together with the IMF used for 47~Tuc modeling in \citet{Giersz47Tuc2011}, which we also adopted here, all have shallower power-law slopes for lower-mass stars than the Kroupa IMF ($\alpha_1 = 1.3\pm0.5$ for $0.08 \leq m/\rm{M_{\odot}} < 0.5$ and $\alpha_2 = 2.3\pm0.3$ for $0.5 \leq m/\rm{M_{\odot}} < 1$; \citealp{Kroupa2001}). In addition, \citet{Baumgardt_Hilker2018} derived a global power-law slope of $0.53$ for $0.2 < m/\rm{M_{\odot}} < 0.8$ for 47 Tuc. This is slightly steeper than the $0.4$ present-day mass function slope for $\sim 0.2 < m/\rm{M_{\odot}} < 0.7$ in our 47~Tuc model, which could be due to the selection bias against faint low-mass stars in the crowded central area of 47~Tuc. In Section~\ref{sec:match} we show that, by adopting an IMF that is both bottom-heavy and top-light, we can produce models that match the 47~Tuc observations. Although we have only explored a small set of initial parameters and did not vary the lower-mass power-law slope of the IMF because the 47 Tuc-size simulations are very computationally expensive. It is possible a power-law slope for lower-mass stars more consistent with that in the Kroupa IMF may be ale to produce a 47~Tuc model.

\subsection{Implications for Black Hole Merger Rate}\label{subsec:bh_merger}
The top-light IMF is likely to reduce the total number of binary BH mergers \citep[e.g.,][]{Weatherford+2021}. Over a Hubble time, the 47~Tuc simulation produced $220$ binary BH mergers, including both those that merged in the cluster and those that merged after being ejected. To check how this steep power-law slope for massive stars affects binary BH mergers, we also calculate the number of mergers in the simulation with high-mass power-law slope $\alpha_2 = 2.3$ (also see Figure~\ref{fig:imf2.3}). Surprisingly, this simulation produces $\sim 2700$ BHs at late times, about $15$ times more than in the best-fitting 47~Tuc models, but only $210$ merging binary BHs over a Hubble time. This occurs because, although it produces more BHs, the model with $\alpha_2 = 2.3$ is much less dense in the core than what we see in the best-fitting 47~Tuc simulation (Figure~\ref{fig:imf2.3}). Thus its BHs undergo fewer dynamical encounters and there are fewer mergers per BH \citep[e.g.,][Figure~13]{Kremer+2020catalog}. Hypothetically, however, if we were to generate 47~Tuc models with a standard Kroupa IMF, but with smaller initial virial radius and a high central density at present, more binary BH mergers should be produced \citep{Weatherford+2021}. In \citet{Weatherford+2021}, models with a high-mass power-law slope $\alpha = 2.3$ produce more binary BH mergers than those with $\alpha = 3$. The difference may be attributed to the very different IMF, initial virial radius and/or initial density profile used in this study than in \citet{Weatherford+2021}.

\subsection{Binary Fraction}\label{subsec:bfrac}
Binary stars play an important role in the dynamical evolution of GCs \citep[e.g.,][]{Heggie_Hut}. For example, BH binaries can act as energy sources in a cluster core, delaying the core collapse of clusters \citep[e.g.,][]{Kremer+2019size}.  Learning the primordial binary fractions of GCs can also help reveal details of the formation of GCs and the properties of their pre-GC molecular clouds. Furthermore, since the majority of stars in the Milky Way galaxy are expected to be born in clusters, the binary fractions in GCs are also connected to that in the field \citep{lada_Lada_2003}. However, the GC binary fractions, especially the relation between primordial and present-day binary fractions, are not well constrained \citep[e.g.,][and references therein]{Fregeau+2009,Milone+2012}. 

\citet{Milone+2012} found the total fraction of binaries in 47~Tuc to be $\sim 2\%$, and the binary fractions at mass ratio $q>0.5, 0.6, 0.7$ to be $0.009\pm0.003, 0.007\pm0.003, 0.005\pm0.003$, respectively at the radial region between the cluster's observed core and half-light radii \citep[][2010 edition]{Harris_1996}. \citet{Ji_Bregman_2015} obtained a binary fraction in 47~Tuc to be $3.01\pm0.13\%$ within about the half-light radius using their direct couting method for binaries with high mass ratios $q > 0.5$. More recently, \citet{Mann+2019} found the binary fractions in 47~Tuc for $q$ in the ranges of $0.5-0.7$, $0.7-0.9$, and $> 0.9$ to be $0.0265$, $0.0098$ and $0.0039$, respectively. These observations are all consistent with low binary fractions.

We ``observed" the binary fractions in the best-fit model by identifying binary stars in its CMD using the technique in \citet{Rui+matching2021}, which mimics the observational cuts in \citet{Milone+2012}. We found that at radii between the observed core and half-light radii \citep[][2010 edition]{Harris_1996}, the best-fit model has binary fractions $0.0107\pm0.00023$, $0.0085\pm0.00021$ and $0.0063\pm0.00018$ at mass ratio $q>0.5, 0.6$ and $0.7$, respectively. This is consistent with the observation in \citet{Milone+2012} at all three mass ratios. 

\begin{figure}
\begin{center}
\includegraphics[width=\columnwidth]{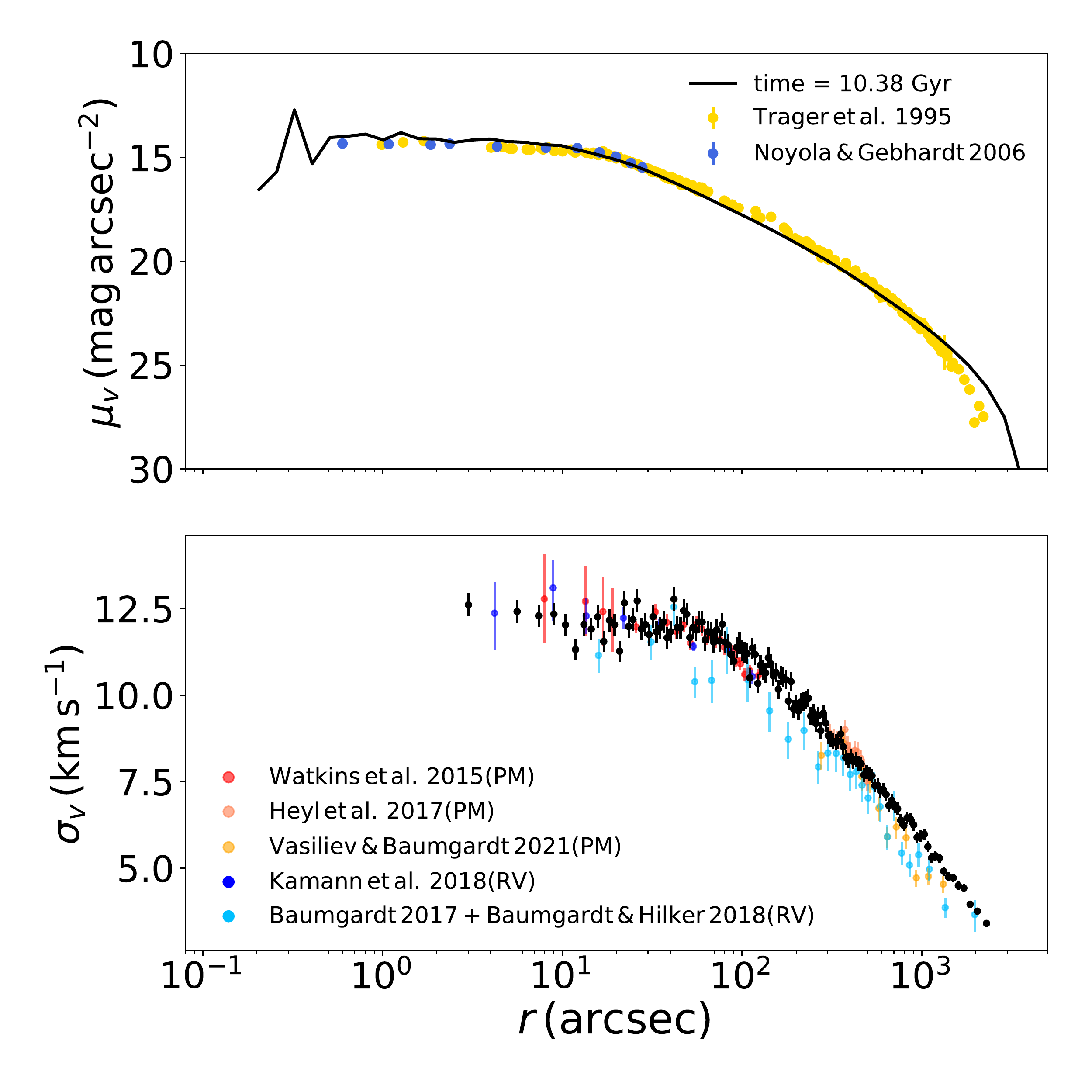}
\caption{A massive model at $t = 10.38~$Gyr with initial binary fraction $f_b = 8\%$ (the black curve and black markers). It has initial $N = 2.95e6$ and initial $r_v = 3.5~$pc. See text for more details. \label{fig:fb10}}
\end{center}
\end{figure}

To explore how changing the primordial binary fraction affects the model, we also experimented with a cluster with initial binary fraction $f_b = 8\%$, initial number of stars $N = 2.95e6$ and initial virial radius $r_v = 3.5~$pc. The small variation in the initial $N$ and $r_v$ is to compensate for the increased number of binaries so that the cluster has similar initial mass to the best-fit 47~Tuc model and that its SBP and VDP at late times will fit the observations. Figure~\ref{fig:fb10} compares the SBP and VDP of this model to the observations, and shows the model properties match well the observations. However, the model has binary fractions $0.0352\pm0.00043$, $0.0274\pm0.00038$, and $0.0199\pm0.0032$ at $q>0.5$, $0.6$, and $0.7$, respectively, which are about three times higher than observed in \citet{Milone+2012}. Furthermore, models with $8\%$ binary fraction may be overproducing MSPs and CVs (there are $132$ MSPs and $787$ CVs at $10.38~$Gyr). Therefore, 47~Tuc is more likely to have a low primordial binary fraction of $< 8\%$ and closer to $\sim 2\%$.

\subsection{Rotation in 47 Tuc}\label{subsec:caveats}
Many studies have shown that 47~Tuc is rotating and has velocity anisotropy using observations from HST, Gaia, and MUSE \citep[e.g.,][]{Watkins+2015,Bellini_2017,Kamann+2018,Vasiliev+2021}. These additional dimensions of the cluster's dynamical structure can improve our understanding of how the cluster forms and evolves, and may alter the cluster properties derived from the traditional fitting of SBPs with spherically symmetric profiles such as the King profile \citep{King1966} or the Elson profile \citep{Elson1987}. However, \texttt{CMC} assumes spherical symmetry \citep{Rodriguez+2021CMC} and is not able to treat rotation. Our model cluster's velocity dispersion is also in general isotropic \footnote{\texttt{CMC} can model clusters with anisotropic velocities \citep{Rodriguez+2021CMC}, but we do not consider anisotropy in this study.}(see Figure~\ref{fig:sigmatr}). Compared to the structural parameters derived in \citet{Bellini_2017} from their best-fit model that took into account rotation and asymmetry, our model has a larger mass but smaller observational core and half-light radii, which are more consistent with the radii in \citet[][2010 edition]{Harris_1996}. Although note that their model SBP \citep[][Figure~9]{Bellini_2017} is slightly lower in the core compared to \citet{Trager1995}, which may contribute to the differences. Given our best-fit model matches the observations using a spherically symmetric Elson profile, the differences in the properties of 47~Tuc when taking into account rotation and velocity anisotropy are likely small.

\section{Conclusions}\label{sec:conclusion}
In this study, we simulate 47~Tuc using the Monte Carlo $N$-body code \texttt{CMC}. As one of the most massive, densest and closest GCs in the Milky Way, there is a large data set of different 47~Tuc observations, but few $N$-body simulations because simulations with the cluster's mass and density are computationally time consuming. We produce 47~Tuc models that can simultaneously match the observed surface brightness, velocity dispersion and number density profiles, the pulsar acceleration profile, and the pulsar distribution in the cluster. These models are self-consistent and take into account all the relevant physics (e.g., BH dynamics) for the first time. The best-fit model has an IMF that is both bottom-heavy and top-light with power-law slopes $\alpha_1 = 0.4$ and $\alpha_2 = 2.8$ (Table~\ref{tab:IC}). The models were also overfilling the tidal radius initially.

We report the number of exotic objects in the models that most closely match the observations; the best-fit models result in $54$ MSPs, $14$ LMXBs, $334$ CVs, $186$ BHs, $1368$ NSs and $101$ BSSs in the cluster at the present day. The numbers of MSPs, LMXBs and CVs are likely upper limits given that our simple treatments of tidal capture and giant star collision to maximize binary formation, and that we do not consider duty cycles and luminosities of LMXBs and CVs. We also demonstrate that tidal capture in GCs may be important to the formation of redback MSPs and many MSPs may be formed from giant star collisions. Furthermore, an intermediate-mass black hole is not necessary to explain the pulsar accelerations observed in the cluster.

We also discuss how different density distribution profiles (King vs. Elson profiles), IMFs, and binary fractions affect the observed properties of 47~Tuc, such as the binary BH merger rates. We point out that there may be a degeneracy in 47~Tuc models given a top-light IMF or an initial small virial radius, which have similar effects on reducing the number of BHs and thus the properties of the cluster. Due to the significant computational time taken to run a 47~Tuc-size simulation, we did not explore systematically the vast parameter space of initial conditions for star clusters, or the full range of uncertainties in our various assumptions and input parameters (e.g., characterizing the BH and NS natal kicks, or the treatment of binary mass transfer and common envelope phases). Therefore, we do not claim that our best-fit model presented here is unique.

\acknowledgments
We thank Sebastian Kamann, Stefan Dreizler, and the anonymous referee for helpful discussions and suggestions on the manuscript. This work was supported by NSF Grants AST-1716762 and AST-2108624 at Northwestern University, and by NSF Grant AST-2009916 at Carnegie Mellon University. K.K. is supported by an NSF Astronomy and Astrophysics Postdoctoral Fellowship under award AST-2001751.  C.R.~acknowledges support from a New Investigator Research Grant from the Charles E.~Kaufman Foundation.  N.Z.R.\ acknowledges support from the Dominic Orr Graduate Fellowship at Caltech and the National Science Foundation Graduate Research Fellowship under Grant No. DGE‐1745301. N.C.W.\ acknowledges support from the CIERA Riedel Family Graduate Fellowship. S.C. acknowledges support from the Department of Atomic Energy, Government of India, under project no. 12-R\&D-TFR-5.02-0200. G.F.\ and F.A.R.\ acknowledge support from NASA Grant 80NSSC21K1722. This research was supported in part through the computational resources and staff contributions provided for the Quest high performance computing facility at Northwestern University, which is jointly supported by the Office of the Provost, the Office for Research, and Northwestern University Information Technology.

\software{\texttt{CMC} \citep{Joshi_2000,Joshi_2001,Fregeau_2003, fregeau2007monte, Chatterjee_2010,Chatterjee_2013b,Umbreit_2012,Morscher+2015,Rodriguez+2016million,Rodriguez+2021CMC}, \texttt{Fewbody} \citep{fregeau2004stellar}, \texttt{COSMIC} \citep{cosmic}}

\vspace{2cm} % Added to prevent text overlap

\appendix
\section{Single-single Interactions in \texttt{CMC}}\label{app:ssinteract}
\begin{figure*}[h]
\begin{center}
\includegraphics[width=\textwidth]{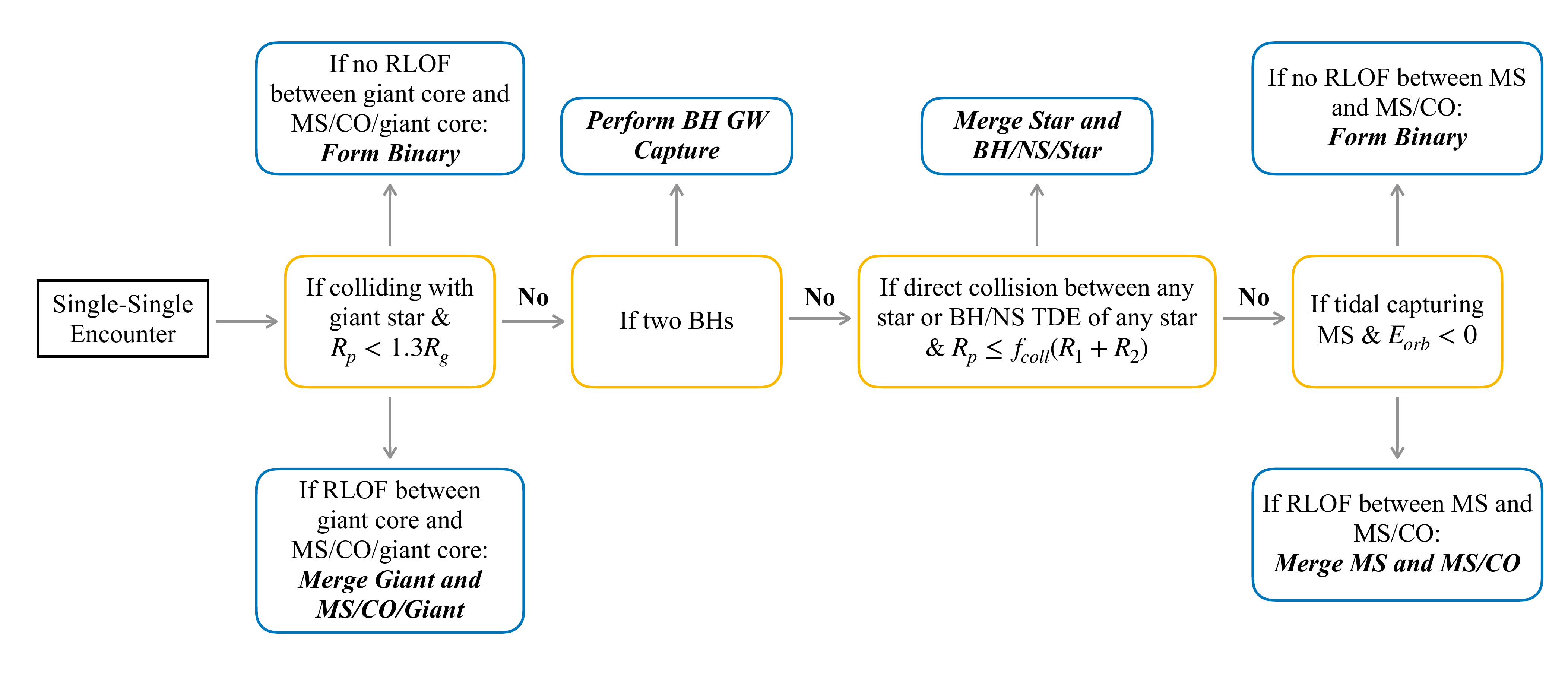}
\caption{Flowchart illustrating single-single dynamical encounters in \texttt{CMC}. Note that if both stars are giants, the collision is performed when $r_p < 1.3 (R_{g1}+R_{g2})$. Default $f_{coll} = 1$ in \texttt{CMC}. ``MS" is main-sequence stars, ``CO" is compact objects, ``RLOF" is Roche lobe overflow and ``TDE" is tidal disruption events. \label{fig:flowchart}}
\end{center}
\end{figure*}

Figure~\ref{fig:flowchart} demonstrates the treatment of all single-single interactions including tidal captures and giant star collisions in \texttt{CMC}.

\section{Comparison with Detailed Stellar Models}\label{app:tccomparison}
\begin{figure*}
\begin{center}
\includegraphics[width=\textwidth]{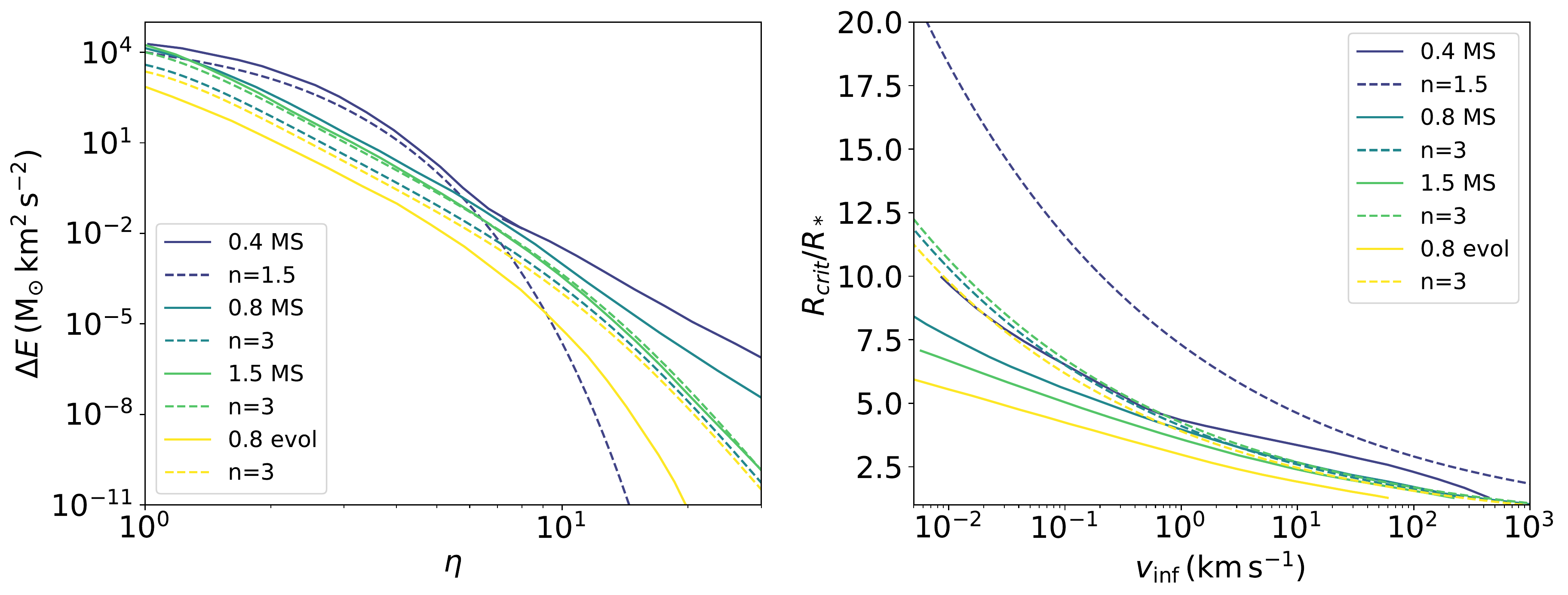}
\caption{Comparison between polytropic stellar models to detailed stellar models of \citet{McMillan+1987}. The left panel shows the tidal energy dissipation in terms of the dimensionless transit time. The right panel shows the ratio of the critical capture periastron separation to the radius of the star as a function of the relative velocity at infinity. ``MS" is for MS stars and ``evol" is for stars just evolve off the main-sequence.  \label{fig:mcmillan}
}
\end{center}
\end{figure*}

To verify that our simple application of polytropic stellar models agrees with more sophisticated calculations, we compare the oscillation energy and the critical capture pericenter distance between the tidal capture of two MS stars in this study with those in \citet{McMillan+1987}. Figure~\ref{fig:mcmillan} shows the comparison. For $\eta \lesssim 10$, the oscillation energy calculated with polytropic stellar models agrees well with the more detailed stellar models. At $v_{\infty} \sim 10~\rm{km\,s^{-1}}$, the velocity dispersion of typical GCs and of 47~Tuc, there can be as big as $\sim 20\%$ discrepancy in the critical pericenter distances between different stellar models for low-mass MS stars, which is not likely to affect the outcomes of tidal capture (see Section~\ref{sec:discuss}). For simplicity, we assume that the structure of the stars does not change during tidal capture.

\section{Tidal Capture and Giant Star Collision Uncertainties}\label{app:tcgc_uncer}
We are only considering tidal capture and giant star collision during single-single encounters in this study for simplicity. To estimate how the tidal capture and giant star collision rates will be affected if we include these interactions during binary-single and binary-binary encounters, we calculated the direct collision (``sticky-sphere" collision) rates for all types of stars in the \texttt{CMC} Cluster Catalog models \citep{Kremer+2020catalog}. We found that for MS stars and compact objects, direct collisions occur $\sim 2.5$ times more frequently during binary-single and binary-binary encounters than single-single encounters. However, for giant stars, direct collisions are $\sim 5$ times more frequent in single-single encounters. Thus including giant star collisions in binary interactions is not likely to have a large influence in the overall rates, but the tidal capture rates is likely to increase by at least twice if the interaction is occuring in binary-binary and binary-single encounters. The implementation of tidal capture into \texttt{Fewbody} is currently undergoing \citep{Vickinprep}.

We assume, during common-envelope evolution, $\alpha = 1$ (parameter for scaling the efficiency of transferring orbital energy to the envelope) and $\lambda = 0.5$ (binding energy factor for stellar envelope) in binary evolution in \texttt{COSMIC} \citep{hurley2002evolution,cosmic}. These simple values give us a sense of how giant star collisions vary with different values for $\alpha$ and $\lambda$. For example, for a smaller binding energy factor where $\lambda$ is close to zero, the release of recombination energy that can enhance envelope ejection is turned off \citep{Claeys+2014} and all giant star collisions lead to mergers. We point out that the values of $\alpha$ and $\lambda$ are highly uncertain \citep[e.g.,][and references therein]{Ivanova_2013_ce}. Previous studies suggested an $\alpha$ from as small as $\sim0.2$ \citep{Zorotovic_2010,Toonen_Nelemans_2013,Camacho_2014}, to as large as $\sim5$ \citep{Fragos_2019}. In addition, while most treatments adopt a constant $\lambda$, it is more likely that $\lambda$ varies according to the types of stars and the stellar mass fraction in their envelopes \citep{Ivanova_2011}. By varying $\alpha$ and $\lambda$ and matching the number and properties of compact objects in GCs, especially pulsars, we could put better constraints on these two parameters but is out of the scale of this study.

Previous studies have shown that tidal-force induced oscillation during tidal capture may cause large fluctuation and instability in MS stars, and may lead to their disruption \citep[e.g.,][]{Kochanek_1992}. We compare the binding energy of the MS stars $0.15\frac{GM^2}{R}$ with their induced oscillation energy during tidal capture \citep{Kochanek_1992}. Only four stars are disrupted for all $R_p$, and none is disrupted when only considering $\frac{R_p}{R}<1.3(\frac{M_T}{2M})^\frac{1}{3}$ \citep{Kochanek_1992}, where $M$, $R$ is the mass and radius of the MS stars, respectively, and $M_T$ is the total mass of the two interacting objects.

Figure~\ref{fig:mcmillan} shows some discrepancies in the oscillation energy and critical radius calculated with polytropic stellar models and detailed stellar models \citep{McMillan+1987}. For the oscillation energy induced by tidal force (left panel), the most pronounced difference is for $n=1.5$ and $0.4~$MS when $\eta \gtrsim 6$. In our 47~Tuc simulation, only about $4\%$ of tidal capture encounters have $\eta > 6$, so the difference for $\eta > 6$ has negligible effect on the final tidal capture binaries. For the critical radius (right panel), the largest deviation is $\sim 20\%$ again for $n=1.5$ and $0.4~$MS at $v_{inf} \sim 10~\rm{km\,s^{-1}}$. However, these parameters mostly involve interactions between low-mass MS stars with another low-mass MS stars or WDs, and the discrepancy will not affect the number of tidal capture NS--MS star binaries. 

\section{All 47~Tuc-Size Simulations}\label{app:allsimulations}
%\begin{longrotatetable}
\begin{deluxetable*}{c|cccccccccccccccc}[h!]
\tabletypesize{\scriptsize}
\tablewidth{0pt}
\tablecaption{Initial Conditions of All Simulations\label{tab:allsim}}
\tablehead{
\colhead{$\rm{No.}$} & \colhead{$\rm{N_{init}}$} & \colhead{$\rm{R_v}$} & \colhead{$\rm{\gamma/W_0}$} & \colhead{$\rm{R_{max}}$} & \colhead{$\rm{Z}$} & \colhead{$\rm{R_{tidal}}$} & \colhead{$\rm{f_b}$} & \colhead{$\rm{Profile}$} & \colhead{$\rm{M_{min}}$} & \colhead{$\rm{\alpha_1}$} & \colhead{$\rm{M_{br1}}$} & \colhead{$\rm{\alpha_2}$} & \colhead{$\rm{M_{br2}}$} & \colhead{$\rm{\alpha_3}$} & \colhead{$\rm{M_{max}}$} & \colhead{$\rm{M_{init}}$}\\
\colhead{} & \colhead{$\rm{10^6}$} & \colhead{$\rm{pc}$} & \colhead{} & \colhead{$\rm{R_v}$} & \colhead{} & \colhead{$\rm{pc}$} & \colhead{} & \colhead{} & \colhead{$\rm{M_{\odot}}$} & \colhead{} & \colhead{$\rm{M_{\odot}}$} & \colhead{} & \colhead{$\rm{M_{\odot}}$} & \colhead{} & \colhead{$\rm{M_{\odot}}$} & \colhead{$\rm{10^6\,M_{\odot}}$}
}
\startdata
$1^*$ & 3.00 & 4.0 & 2.1 & 200 & 0.0038 & 182.17 & 0.022 & E & 0.08 & 0.4 & 0.8 & 2.8 & -- & -- & 150 & 2.48\\
2 & 3.00 & 4.0 & 2.1 & 200 & 0.0038 & 149 & 0.022 & E & 0.08 & 0.4 & 0.8 & 2.8 & -- & -- & 150 & 2.52\\
3 & 2.95 & 3.5 & 2.1 & 200 & 0.0038 & 171 & 0.08 & E & 0.08 & 0.4 & 0.8 & 2.8 & -- & -- & 150 & 2.51\\
4 & 0.80 & 4.0 & 2.1 & 200 & 0.0038 & 182.17 & 0.022 & E & 0.08 & 0.4 & 0.8 & 1.8 & -- & -- & 150 & 2.57\\
5 & 1.23 & 4.0 & 2.1 & 200 & 0.0038	& 182.17 & 0.022 & E & 0.08 & 0.4 & 0.8 & 2 & -- & -- & 150 & 2.58\\
6 & 1.96 & 4.0 & 2.1 & 200 & 0.0038	& 182.17 & 0.022 & E & 0.08 & 0.4 & 0.8 & 2.3 & -- & -- & 150 & 2.56\\
7 & 2.80 & 4.0 & 2.1 & 200 & 0.004 & 171 & 0.022 & E & 0.08	& 0.52 & 0.5 & 1.35 & 1 & 2.49 & 150 & 2.66\\
8 & 2.80 & 3.0 & 2.1 & 200 & 0.004 & 171 & 0.022 & E & 0.08	& 0.52 & 0.5 & 1.35 & 1 & 2.49 & 150 & 2.66\\
9 & 2.00 & 1.4 & 7.5 & -- & 0.0038 & 159.15 & 0.022 & K & 0.08 & 0.4 & 0.8 & 2.8 & -- & -- & 150 & 1.66\\
10 & 2.00 & 1.8 & 7.5 & -- & 0.0038 & 159.15 & 0.022 & K & 0.08 & 0.4 & 0.8 & 2.8 & -- & -- & 150 & 1.66\\
11 & 2.00 & 1.8 & 2.1 & 300 & 0.0038 & 159.15 & 0.022 & E & 0.08 & 0.4 & 0.8 & 2.8 & -- & -- & 150 & 1.66\\
12 & 2.00 & 4 & 2.1 & 300 & 0.0038 & 159.15 & 0.022 & E & 0.08 & 0.4 & 0.8 & 2.8 & -- & -- & 150 & 1.66\\
13 & 2.40 & 4 & 2.1 & 300 & 0.0038 & 169.13 & 0.022 & E & 0.08 & 0.4 & 0.8 & 2.8 & -- & -- & 150 & 1.99\\
14 & 2.00 & 1.84 & 7.5 & -- & 0.0038 & 158.32 & 0.022 & K & 0.08 & 0.4 & 0.8 & 2.8 & -- & -- & 50 & 1.63\\
15 & 2.00 & 1.84 & 7.5 & -- & 0.0038 & 129.4324061 & 0.022 & K & 0.08 & 0.4 & 0.8 & 2.8 & -- & -- & 50 & 1.63\\
\enddata
\tablecomments{Columns from left to right: number of objects, virial radius, power-law slope of an Elson profile or King concentration parameter, maximum radius (in virial radius) to sample the cluster, metallicity, tidal radius, binary fraction, E--Elson or K--King profile, masses and slopes of the IMF, and total cluster mass.\\ *--the best-fit 47~Tuc simulation\\ No. 2 and 3 also fit the SBPs and VDPs, so are used for calculating the more general uncertainties of the properties and numbers shown in Table~\ref{tab:present_prop}.\\ No. 4 and 5 disrupted at early times of the simulation. \\ The run of No. 8 was stopped at early times because it is very computationally expensive. The IMF for No. 7 and 8 are adopted from \citet{Henault-Brunet+2019}.\\ The tidal radius $171$ pc is adopted from \citet{Lane_2012}.}
\end{deluxetable*}
%\end{longrotatetable}

\startlongtable
\begin{deluxetable*}{c|cccccccccccccc}
\tabletypesize{\scriptsize}
\tablewidth{0pt}
\tablecaption{Present-Day Properties of Best- and Near-fit Models \label{tab:general_prop}}
\tablehead{
\colhead{} & \colhead{$\rm{Mass}$} & \colhead{$\rm{r_c}$} & \colhead{$\rm{r_h}$} & \colhead{$\rm{r_{cobs}}$} & \colhead{$\rm{r_{hlobs}}$} & \colhead{$\rm{N_{BH}}$} & \colhead{$\rm{N_{BH-BH}}$} & \colhead{$\rm{N_{NS}}$} & \colhead{$\rm{N_{LMXB}}$} & \colhead{$\rm{N_{MSP}}$} & \colhead{$\rm{N_{young\,psr}}$} & \colhead{$\rm{N_{CV}}$} & \colhead{$\rm{N_{BSS}}$} & \colhead{$\rm{N_{tcNS-MS}}$}\\
\colhead{} & \colhead{$\rm{M_{\odot}}$} & \colhead{$\rm{pc}$} & \colhead{$\rm{pc}$} & \colhead{$\rm{arcmin}$} & \colhead{$\rm{arcmin}$} & \colhead{} & \colhead{} & \colhead{} & \colhead{} & \colhead{} & \colhead{} & \colhead{} & \colhead{} & \colhead{}
}
\startdata
$\mu$ & $9.86\times10^5$ & 0.82 & 6.48 & 0.33 & 2.64 & 160 & 2 & 1369 & 17 & 50 & 0 & 412 & 135 & 5 \\
$\sigma$ & $2.82\times10^4$ & 0.09 & 0.34 & 0.04 & 0.22 & 32 & 1 & 1 & 4 & 3 & 0 & 177 & 72 & 3 \\
\enddata
\tablecomments{Similar to Table~\ref{tab:present_prop}, mean values and standard deviations of cluster properties and numbers of compact objects. The last column shows the average number of tidal capture NS--MS binaries and its standard deviation. We relaxed the $\chi^2$ selection criterion described in Section~\ref{sec:match} to include more snapshots at $\sim 8.3-13.5$ Gyr from the best-fit 47~Tuc simulation, and also snapshots from simulation No. 2 ($\sim 9-13.1$ Gyr) and 3 ($\sim 10-13.6$ Gyr) in Table~\ref{tab:allsim}. Most mean values and standard deviations stay similar to those in Table~\ref{tab:present_prop}. There are large standard deviations for the numbers of CVs and BSSs because simulation No. 3 with a higher initial binary fraction produces many more of these systems.}
\end{deluxetable*}

\bibliographystyle{aasjournal}
\bibliography{47Tuc}

\end{document}